\documentclass[%
 reprint,
 amsmath,amssymb,
 aps,
superscriptaddress,
]{revtex4-2}

\usepackage[dvipsnames]{xcolor}
\usepackage{graphicx}
\usepackage{dcolumn}
\usepackage{bm}
\usepackage{hyperref}
\usepackage{placeins}
\usepackage[mathlines]{lineno}
\usepackage{siunitx}
\usepackage{braket}

\DeclareSIUnit\gauss{G}
\usepackage{tikz}

\definecolor{verdealberto}{rgb}{0.05, 0.64, 0.09}

\begin{document}

\preprint{APS/123-QED}

\title{Optically trapped {F}eshbach molecules of fermionic $^{161}$Dy and $^{40}$K}

\author{E. Soave}
\altaffiliation[Present address: ]
{Bosch Sensortec GmbH, Reutlingen 72770, Germany}
\affiliation{Institut f{\"u}r Experimentalphysik, Universit{\"a}t Innsbruck, Austria} 

\author{A. Canali}
\affiliation{Institut f{\"u}r Experimentalphysik, Universit{\"a}t Innsbruck, Austria} 

\author{Zhu-Xiong Ye}
\affiliation{Institut f{\"u}r Experimentalphysik, Universit{\"a}t Innsbruck, Austria} 

\author{M. Kreyer}
\affiliation{Institut f{\"u}r Experimentalphysik, Universit{\"a}t Innsbruck, Austria} 

\author{E. Kirilov}
\affiliation{Institut f{\"u}r Experimentalphysik, Universit{\"a}t Innsbruck, Austria} 
\affiliation{Institut f{\"u}r Quantenoptik und Quanteninformation (IQOQI), {\"O}sterreichische Akademie der Wissenschaften, Innsbruck, Austria}

\author{R. Grimm}
\affiliation{Institut f{\"u}r Experimentalphysik, Universit{\"a}t Innsbruck, Austria} 
\affiliation{Institut f{\"u}r Quantenoptik und Quanteninformation (IQOQI), {\"O}sterreichische Akademie der Wissenschaften, Innsbruck, Austria}

\date{\today}

\begin{abstract}

We report on the preparation of a pure ultracold sample of bosonic DyK Feshbach molecules, which are composed of the fermionic isotopes $^{161}$Dy and $^{40}$K. Employing a magnetic sweep across a resonance located near 7.3\,G, we produce up to 5000 molecules at a temperature of about 50\,nK. For purification from the remaining atoms, we apply a Stern-Gerlach technique based on magnetic levitation of the molecules in a very weak optical dipole trap. With the trapped molecules we finally reach a high phase-space density of about 0.1. We measure the magnetic field dependence of the molecular binding energy and the magnetic moment, refining our knowledge of the resonance parameters. We also demonstrate a peculiar anisotropic expansion effect observed when the molecules are released from the trap and expand freely in the magnetic levitation field. Moreover, we identify an important lifetime limitation that is imposed by the 1064-nm infrared trap light itself and not by inelastic collisions. The light-induced decay rate is found to be proportional to the trap light intensity and the closed-channel fraction of the Feshbach molecule. These observations suggest a one-photon coupling to electronically excited states to limit the lifetime and point to the prospect of loss suppression by optimizing the wavelength of the trapping light. Our results represent important insights and experimental steps on the way to achieve quantum-degenerate samples of DyK molecules and novel superfluids based on mass-imbalanced fermion mixtures.

\end{abstract}

\maketitle

\section{Introduction}\label{sec:Intro}

Over the past twenty years, the formation of molecules via magnetically tuned Feshbach resonances \cite{Koehler2006poc, Chin2010FRI} has been established as a powerful tool for experiments on ultracold quantum gases, which allows for coupling of the atomic to the much richer molecular world.
Early experiments on Bose-Einstein condensates (BECs) have demonstrated
atom-molecule coupling \cite{Wynar2000mia, Donley2002amc} and the production of molecular samples near quantum degeneracy \cite{Herbig2003poa, Xu2003foq, Duerr2004oom}. In ultracold Fermi gases, molecule formation \cite{Regal2003cum, Strecker2003coa, Cubizolles2003pol, Jochim2003pgo} plays a particular role because of the inherent change of the quantum statistics when pairs of fermions form composite bosons. 
This has led to the observation of molecular BEC \cite{Jochim2003bec, Greiner2003eoa, Zwierlein2003oob}, the demonstration of fermionic condensation \cite{Regal2004oor} and superfluidity \cite{Zwierlein2005vas} and a great wealth of experiments on the crossover in strongly interacting Fermi gases from BEC to Bardeen-Cooper-Schrieffer (BCS) type pairing \cite{Zwerger2012tbb,Strinati2018tbb}.

In heteronuclear molecular systems, which in the laboratory can be realized with various two-species atomic mixtures \cite{Ospelkaus2006uhm, Weber2008aou, Spiegelhalder2010aop, Wu2012uff, Heo2012fou, Koeppinger2014poo, Takekoshi2014uds, Wang2015fou, Lam2022HPS}, the interest in Feshbach resonances has been to large extent driven by the possibility to create polar molecules featuring strong electric dipole-dipole interactions. For mixtures of two fermionic species \cite{Taglieber2008qdt, Wille2008eau, Voigt2009uhf, Naik2011fri, Hara2011qdm, Green2020fri, Ravensbergen2018poa, Ravensbergen2020rif, Ciamei2022ddf, Ciamei2022euc}, however, another promising application is stimulating the interest in Feshbach resonances and molecules. Theoretical work has predicted fermionic quantum-gas mixtures with mass imbalance to favor exotic interaction regimes~\cite{Gubbels2013ifg}. Mass-imbalanced systems hold particular promise in view of superfluid states with unconventional pairing mechanisms~\cite{Gubbels2009lpi, Wang2017eeo, Pini2021bmf}, most notably the elusive Fulde-Ferrell-Larkin-Ovchinnikov (FFLO) state~\cite{Fulde1964sia, Larkin1964nss, Radzihovsky2010ifr}. In addition to the exciting prospects in many-body physics, a variety of interesting few-body phenomena have been predicted to emerge in resonant fermion mixtures \cite{Naidon2017epa, Kartavtsev2007let}.

Our system is an ultracold Fermi-Fermi mixture of $^{161}$Dy and $^{40}$K atoms \cite{Ravensbergen2018poa}, which features a great variety of interspecies Feshbach resonances at high and low magnetic fields \cite{Ravensbergen2020rif, Ye2022OOL}. In our recent work \cite{Ye2022OOL}, we have identified a particularly interesting resonance in the low-field region near 7\,G, which is isolated from other resonances and strong enough to facilitate accurate magnetic interaction tuning in combination with a significant universal range. We have also demonstrated the formation of Feshbach molecules by a magnetic sweep across the resonance. 

In the present article, we report on the optical trapping of our DyK  Feshbach molecules. In Sec.~\ref{Sec:PreparationSection} we show how the molecular cloud is purified by a magnetic levitation technique, where the molecules are kept in the shallow optical dipole trap while the atoms are pulled out of it. In Sec.~\ref{Sec:ResonanceParameter} we present a state-of-the-art characterization of the Feshbach resonance, based on the measurement of molecular binding energies and of the molecular magnetic moment. In Sec.~\ref{Sec:Antitrapping} we discuss an anisotropic expansion effect that is observed when the cloud is released close to the resonance in the presence of the magnetic levitation field. In Sec.~\ref{Sec:Lifetime} we study the lifetime of the trapped molecular cloud and identify a light-induced one-body mechanism (and not inelastic collisions) as the main limiting factor. In Sec.~\ref{Sec:Outro} we summarize the main conclusions of our work and point to future prospects.

\section{Preparation of pure molecular samples}\label{Sec:PreparationSection}

In this section we present the main steps to produce a pure sample of trapped DyK Feshbach molecules. The molecules are associated using the resonance near 7G that we have identified in our recent work  \cite{Ye2022OOL}. In Sec.\,\ref{SubSec:PrepMixture} we briefly summarize how a double-degenerate sample of Dy and K is obtained and how it is transferred into the 7-G magnetic field region. In Sec.\,\ref{SubSec:SGPurification} we present the experimental procedure that results in the formation of a pure molecular sample. The purification is based on a Stern-Gerlach (SG) separation of the molecular cloud, held in a weak optical dipole trap (ODT), from the atomic clouds, which are pulled out of the trap. As discussed in detail in Sec.\,\ref{SubSec:TemperatureAndPSD}, this procedure allows us to obtain a sample of about \SI{5e3}{} DyK molecules at a temperature of about \SI{50}{\nano\kelvin}. We reach a phase-space density (PSD) of about 0.14, only one order of magnitude away from degeneracy.

\subsection{Preparation of the mixture and transfer into the 7-G magnetic field region}\label{SubSec:PrepMixture}

The starting point of every experimental cycle is the production of a degenerate sample of Dy and K at \SI{250}{\milli\gauss} in a near-infrared (\SI{1064}{\nano\meter}) ODT. Both species are spin polarized in their lowest hyperfine sublevels $\ket{F,m_F}=\ket{21/2,-21/2}$ and $\ket{9/2,-9/2}$, respectively. Following the procedures already presented in Refs.\ \cite{Ravensbergen2018poa}  and \cite{Ye2022OOL}, we are able to reach the deeply degenerate regime with reduced temperatures $T_{\mathrm{Dy}}/T_F^{\mathrm{Dy}}\approx0.13$ and $T_{\mathrm{K}}/T_F^{\mathrm{K}}\approx0.13$, normalized to the respective Fermi temperatures.
The sample is then transferred into a \SI{1064}{\nano\meter} crossed ODT formed by a horizontal beam, with a waist of \SI{100}{\micro\meter} (Azurlight ALS-IR-1064-5-
I-CC-SF) and a vertical beam, waist of \SI{65}{\micro\meter} (Mephisto MOPA 18 NE). A magnetic field gradient of \SI{2.2}{\gauss/\centi\meter} supports the trapping of Dy, without fully canceling  the differential gravitational sag with respect to K. This guarantees that the two clouds are kept spatially separated throughout the subsequent phase, during which the magnetic field is ramped within \SI{2}{\milli\second} to $B=\SI{7.775}{\gauss}$, corresponding to a \SI{500}{\milli\gauss} detuning above resonance.

To address specific magnetic field strengths in the 7-G region, we employ a set of coils (offset coils) that provide a fixed stable offset, which remains unchanged for the rest of the sequence. These coils were originally designed to produce high magnetic fields of a few hundred gauss \cite{Ravensbergen2020rif} and respond rather slowly to magnetic field changes. Therefore, the magnetic field ramps required for the rest of the sequence are performed employing an additional set of coils ({fast coils}), whose magnetic field adds to the constant field provided by the offset coils. Further details on the way we control the magnetic field during the experimental sequence are reported in App.\,\ref{App:MagneticFieldControl}.

\subsection{Molecule association and Stern-Gerlach purification}\label{SubSec:SGPurification}
\begin{figure}
\centering
\includegraphics[
width=1\columnwidth
]{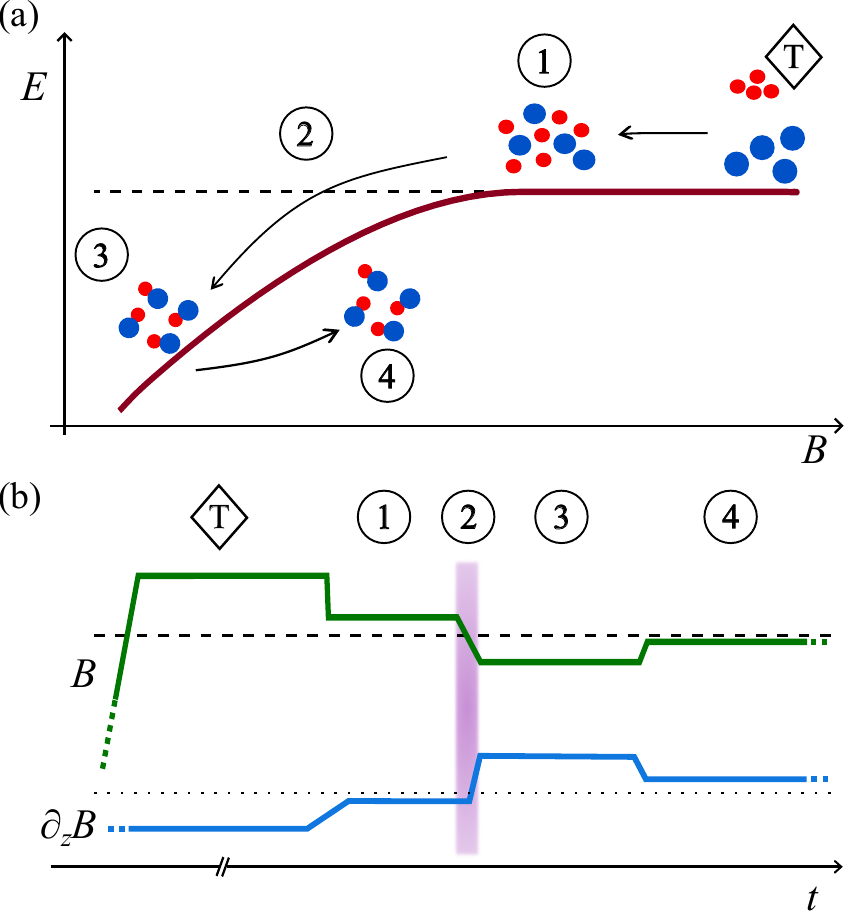}
\caption{Procedure to prepare a pure sample of Feshbach molecules by magnetic field ramping. (a) Qualitative representation of the steps of the association process and the SG purification in relation to the magnetic field and the energy state of free atoms and molecules. The dashed line refers to the unbound atomic pair below resonance. (b) Qualitative representation of the corresponding ramping scheme for the magnetic field (green) and magnetic field gradient (blue); see main text for details. The dashed line represents the resonance pole position and the dotted line represents the magnetic field gradient that levitates dysprosium. After the transfer of the sample into the 7-G region (T, diamond symbol), the sequence proceeds with four main steps: 1) overlap of the two clouds; 2) magnetic field association of molecules (pink shaded region); 3) SG purification of the sample; 4) transfer to the desired magnetic field. }\label{Fig1:Sequence}
\end{figure}
Figure\,\ref{Fig1:Sequence} schematically illustrates the part of the experimental sequence that takes place in the 7-G region.
At the end of the transfer into the 7-G region, the two clouds are kept spatially separate for about \SI{50}{\milli\second} to let the field fully stabilize. The experimental sequence then proceeds in four further steps: (1) spatially overlapping the two clouds, (2) magnetic association of molecules, (3) SG purification of the molecular sample, and (4) adiabatic transfer to the desired target field and preparation for final experiments. The molecular state adiabatically follows the changes in magnetic field. In order to account for the changes in the molecular magnetic moment, the magnetic field ramps performed in the steps (1) to (4) are accompanied by corresponding magnetic field gradient ramps, this ensures that the molecular sample experiences the same magnetic force.

We now discuss the four stages in more detail. 

Stage (1): We apply a 12-ms ramp of the magnetic field gradient from its initial value to the one that cancels the differential gravitational sag between Dy and K so that the two clouds overlap. In the following this is referred to as ``magic levitation'' \cite{Lous2017toa},  realized at $\partial_z B=\SI{2.69}{\gauss/\centi\meter}$. Simultaneously with the gradient ramp, the trapping frequencies are reduced from $\bar{\omega}_{\mathrm{Dy}}=2\pi\times\SI{50}{\hertz}$ to about $2\pi\times\SI{35}{\hertz}$, where $\bar\omega$ is the geometrically averaged trap frequency ($\omega_{\rm K}\approx 3.6\,\omega_{\rm Dy}$ \cite{Ravensbergen2018ado}).
Half-way during the ramp, the magnetic field is decreased within \SI{1}{\milli\second} to \SI{7.325}{\gauss}, about \SI{+50}{\milli\gauss} above the resonance center. 

Stage (2): By magnetic field association \cite{Koehler2006poc, Chin2010FRI} we create about \SI{6e3}{} DyK molecules, corresponding to almost 30\%  conversion efficiency. We found the optimal ramp speed to be $\dot{B}=\SI{0.43}{\gauss/\milli\second}$, which is in the range of typical experiments using Feshbach resonances of similar widths \cite{Chin2010FRI}. The 0.4-ms ramp ends at a magnetic detuning of about \SI{-120}{\milli\gauss}, which corresponds to a molecular binding energy of $h\times$\SI{220}{\kilo\hertz}. We identify such a magnetic field as optimal for the purification of the molecular sample in the subsequent stage (3). The second half of the magnetic field sweep are accompanied by an increase of the magnetic field gradient from the magic levitation value to the one required to balance the gravitational force on the molecules at this detuning. For the sake of notation simplicity, we introduce the {molecular levitation gradient} ${B'}_{\rm{mol}}=m_{\mathrm{mol}}g/\mu_{\mathrm{mol}}$, where $m_{\mathrm{mol}}$ is the molecular mass, $g$ is the gravitational acceleration and $\mu_{\mathrm{mol}}$ is the molecular magnetic moment. Note that $\mu_{\rm mol}$ and thus ${B'}_{\rm{mol}}$ are $B$-dependent quantities. At the end of the magnetic field sweep, the sample consists of a three-component mixture made of DyK molecules and unbound Dy and K atoms.
\begin{figure}
\centering
\includegraphics[
width=1\columnwidth
]{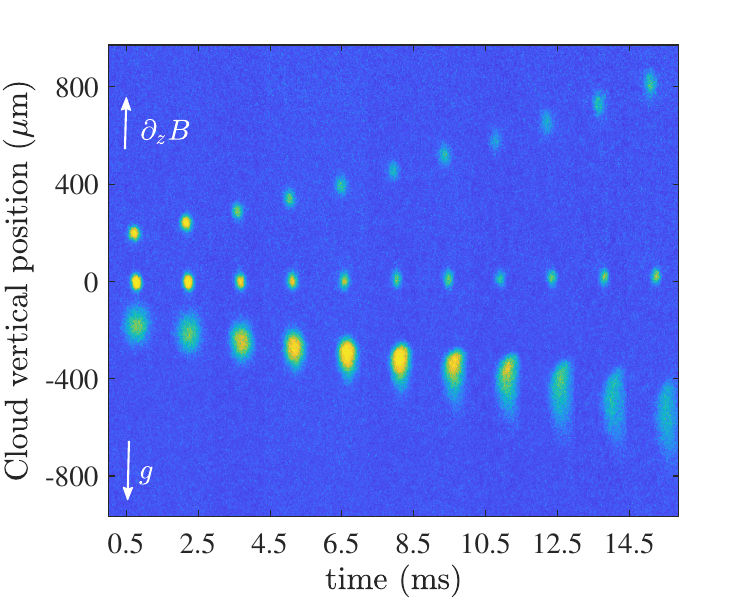}
\caption{Demonstration of SG purification. After magnetic association, the molecules are held in an very shallow trap, at $B=\SI{7.170}{\gauss}$. The magnetic field gradient is set to cancel the gravitational force experienced by the molecules, which therefore are kept in the ODT (central cloud, held at a fixed position $z=0$). Dysprosium (at cloud positions $z>0$) and potassium (at cloud positions $z<0$) undergo a vertical acceleration. Each image is the average of four absorption pictures, taken after a variable hold time (plus a fixed expansion time of {\SI{4}{\milli\second}}). The molecular signal corresponds to the dissociated potassium component of the molecular cloud, and has been imaged together with the pure K atoms. The absorption pictures taken of K and Dy have been superposed for presentation purposes.}\label{Fig2:Spilling}
\end{figure}

Stage (3): We proceed with the purification of the molecular sample. The depth of the ODT is decreased to about $U_{\rm DyK}=k_B\times\SI{220}{\nano\kelvin}$, so that only the molecules, perfectly levitated at ${B'}_{\rm {mol}}$, are kept in the trap. This naturally leads to a SG separation of the three components. Figure \ref{Fig2:Spilling} demonstrates the SG separation process, showing a series of absorption images of the three clouds, taken at different hold times. The end of the magnetic field association sweep corresponds to $t=0$. The molecular cloud remains at the same vertical position ($z=0$) and does not expand, confirming that the molecules are indeed kept in the trap. Dy and K are instead accelerated vertically, being over- and under-levitated, respectively.  In the figure the motion appears as a parabolic trajectory with positive (Dy) and negative (K) curvature. The magnetic field for the purification is chosen in order to maximize the difference in accelerations between the three clouds (Sec.\,\ref{SubSec:MagnMomentSpect}), while minimizing molecular losses (Sec.\,\ref{Sec:Lifetime}).
After \SI{4}{\milli\second} the atomic clouds are far enough from the ODT that they can no longer be recaptured. At this point we are left with a pure molecular sample. We note that a similar purification scheme, based on magnetic levitation and SG separation, was demonstrated for optically trapped RbCs Feshbach molecules in Ref.\,\cite{Koeppinger2014poo}.

Stage (4): The magnetic field strength $B$ is finally ramped to its target value within typically 0.5 ms, and the magnetic field gradient is varied accordingly to follow $B'_{\rm{mol}}$. The trapping conditions (trap frequencies, depth, geometry) are modified depending on the specific measurement we want to perform with the pure sample of DyK Feshbach molecules in the last stage of the experimental cycle. 

At the end of the sequence, the ODT is switched off. The molecular cloud is let to expand keeping the magnetic field and magnetic field gradient fixed. One ms before the imaging pulse, the magnetic field is ramped to positive detunings of the resonance and the molecules are dissociated. The main information about the molecules is obtained from the absorption picture of the dissociated potassium component of the molecular cloud. We can also image the molecules without dissociating them, while keeping the magnetic field fixed. This possibility has the advantage of not altering the expansion of the molecules in time of flight, but the efficiency of the absorption imaging depends on the magnetic field detuning as for more deeply bounded molecular state (large negative $B$ field detunings) only a fraction of the total molecule number can be imaged.
\subsection{Temperature and phase-space density}\label{SubSec:TemperatureAndPSD}
\begin{figure}
\centering
\includegraphics[
width=1\columnwidth
]{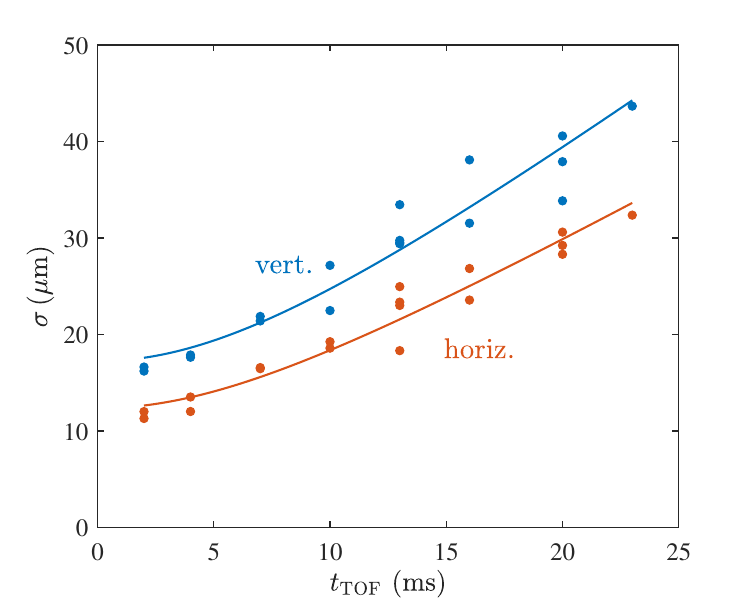}
\caption{Vertical and horizontal size of the molecular cloud as a function of the time of flight $t_{\rm TOF}$. The sample is let expand at $B=\SI{7.22}{\gauss}$, upon switching off the magnetic levitation together with the ODT. A fit to the data, assuming a ballistic expansion, yields a vertical and horizontal temperature of \SI{77(6)}{} and \SI{45(3)}{\nano\kelvin}, respectively.
}\label{Fig3:Temperature}
\end{figure}
The experimental scheme described above provides us with a pure molecular sample of about $N_{\rm{mol}}=\SI{5e3}{}$ molecules, at a temperature of few tens of \SI{}{\nano\kelvin}. We derive the temperature of the molecular cloud from a time-of-flight expansion (Fig.\,\ref{Fig3:Temperature} \footnote{See Supplemental Material for the data files of all figures}). In this specific set of experiments, the molecules are not dissociated to avoid extra energy from the dissociation, but imaged directly \cite{Duerr2004dou}.  A fit to the data reported in Fig.~\ref{Fig3:Temperature} yields a horizontal and a vertical temperature of \SI{45(3)}{\nano\kelvin} and \SI{77(6)}{\nano\kelvin}, respectively. We attribute the difference in these results to additional excitations caused by the rapid changes of vertical magnetic forces during the ramping process. We indeed observe uncontrolled vertical collective oscillations affecting the expansion, which can result in increased temperatures if measured vertically. The horizontal behavior is free of such effects and the expansion can be fully attributed to the temperature of the sample.

We are able to directly measure the trap frequencies of the molecules by observing the center of mass oscillations. Assuming typical trapping conditions ($\bar{\omega}=2\pi\times\SI{28}{\hertz}$), we calculate a phase-space density $\mathrm{PSD}=N_{\mathrm{mol}}(\hbar\bar{\omega}/k_BT_{\mathrm{mol}})^3\approx0.14$, about one order of magnitude away from degeneracy. The conditions near quantum degeneracy represent an excellent starting point for further experiments. Comparable values of the PSD have been reached recently in a system of NaCs Bose-Bose Feshbach molecules \cite{Lam2022HPS}. While BEC has not yet been reached with bosonic heteronuclear Feshbach samples, quantum degenracy has been obtained in fermionic heteronuclear systems of KRb\cite{Demarco2019adf} and NaK \cite{Duda2023tfa}.


\section{Binding energy, magnetic moment, and resonance parameters}
\label{Sec:ResonanceParameter}
In this section, we provide a further characterization of the Feshbach resonance near \SI{7}{\gauss}, already introduced in our previous work \cite{Ye2022OOL}. We first, in Sec.\,\ref{SubSec:BasicDefinition}, introduce the definitions of the relevant parameters and some basic relations. We then, in Secs.\,\ref{SubSec:BindingEnergyMeas} and \ref{SubSec:MagnMomentSpect}, report on two independent sets of experiments, based on measurements of the magnetic-field dependent molecular binding energy and the magnetic moment, which further confirm and refine the values of the key parameters describing the resonance. In Sec.\,\ref{SubSec:FBParametersDiscussion}, we summarize our results. 

\subsection{Basic definitions and relations}\label{SubSec:BasicDefinition}
Close to the center of a single isolated $s$-wave Feshbach resonance, the scattering length $a$ is large and can be approximated by
\begin{equation}
a= - \frac{A}{\delta B} a_0 ,
\end{equation}
where $A$ is a parameter characterizing the resonance strength, $\delta B = B-B_0$ is the magnetic detuning from the resonance center $B_0$, and $a_0$ denotes the Bohr radius.
For a narrow Feshbach resonance \cite{Chin2010FRI}, a proper characterization of the system also requires the knowledge of a third parameter, such as the difference in magnetic moments of the open channel (atom pair) and closed channel (uncoupled molecular state), which we define as $\delta\mu=\mu_{\mathrm{open}}-\mu_\mathrm{closed}$.
For describing narrow Feshbach resonances it is convenient to introduce a characteristic length scale, the range parameter $R^*=\hbar^2/(2m_ra_0\delta\mu A)$, with $m_r$ being the reduced mass of the atom pair \cite{Petrov2004TBP}. The condition $a\gtrsim R^*$ characterizes the crossover into the universal regime, where the two-, few- and many-body interaction physics acquires universal properties \cite{Chin2010FRI, Braaten2006UIF, Bloch2008MBP}. 
For convenience, we introduce a corresponding magnetic-field scale $\delta B^*$, defined as the negative detuning where $a(-\delta B^*)=4R^*$. In terms of the resonance parameters $A$ and $\delta\mu$, this corresponds to
\begin{equation}
\delta B^* =m_r\,\delta\mu\,a_0^2\,A^2/(2\hbar^2).
\end{equation}
In our case, a negative detuning gives a positive scattering length and corresponds to the magnetic field region where the pair can bind into a molecule. Knowing $B_0, \delta B^*$ and $\delta\mu$, the energy of the molecular state relative to the atomic pair state (negative binding energy)  can be calculated as \cite{Lous2018PTI, Petrov2004TBP}
\begin{equation}\label{Eq:BindingEnergy}
    E_{\mathrm{mol}}=-\delta\mu\,\delta B^*\left(\sqrt{1-\frac{\delta B}{\delta B^*}}-1 \right)^2.
\end{equation}
The differential magnetic moment of the molecules is related to the binding energy via the differential relation
\begin{equation}\label{Eq:MagneticMoment}
    \delta\mu_{\mathrm{mol}}=\partial E_{\mathrm{mol}}/\partial B=\delta\mu\left(1-\frac{1}{\sqrt{1-\frac{\delta B}{\delta B^*}}}\right).
\end{equation}

In the following two subsections, we report on measurements of the magnetic-field dependence of the molecular binding energy and the differential magnetic moment. 
\subsection{Binding energy spectroscopy}\label{SubSec:BindingEnergyMeas}
\begin{figure}
\centering
\includegraphics[
width=1\columnwidth
]{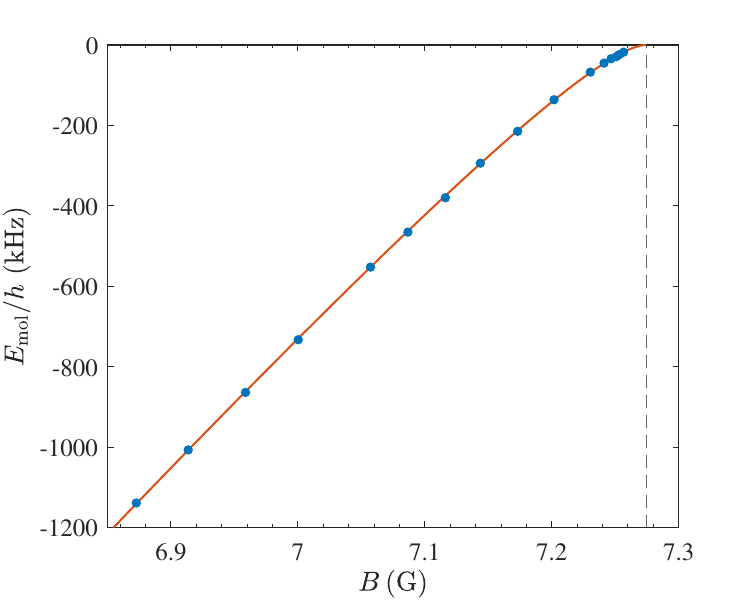}
\caption{
Molecular state energy as a function of the magnetic field, measured via modulation of the magnetic field (wiggle spectroscopy). The experimental data have been corrected by the small thermal shift of $\sim\SI{5}{\kilo\hertz}$. Uncertainties are smaller than the symbol size. The solid line is a fit to the data according to Eq.\,(\ref{Eq:BindingEnergy}). The dashed line indicates the position of the resonance pole.
}\label{Fig4:Wiggle}
\end{figure}
A common method to determine the resonance parameters is based on measuring the molecular binding energy as a function of the magnetic field by magnetic-field (wiggle) modulation spectroscopy \cite{Thompson2005UMP, Donley2002amc, Claussen2003VHP}.

The mixture is prepared at low magnetic-field strength and transferred in the 7-G region, as described in Sec.\,\ref{SubSec:PrepMixture}. By means of the fast coils, the field is then quickly ramped to the target field below the resonance center. We finally superpose the two clouds, ramping the magnetic field gradient to reach magic levitation conditions. At this point, our sample consists  of {$N_{\mathrm{K}}=\SI{1.5e4}{}$} atoms of potassium and {$N_{\mathrm{Dy}}=\SI{7e4}{}$} atoms of dysprosium at a temperature of about \SI{160}{\nano\kelvin} (for this measurement we do not perform full evaporation).
We associate the molecules by a sinusoidal modulation of the magnetic field around the bias field. The modulated field is produced by a coil with four loops, which is placed on the top viewport, tilted with respect to the magnetic field, and driven by a radio-frequency (RF) signal generator. Taking into account the relative kinetic energy $E_{\mathrm{kin}}$ of the thermal motion of an atom pair, the association of a dimer happens when the resonance condition $hf_{\mathrm{mod}}=-E_{\mathrm{mol}}+E_{\mathrm{kin}}$ is fulfilled \cite{Hanna2007AOM,Weber2008aou,Mohapatra2015HAS}, and manifests itself in atomic loss. With a cloud temperature $T$ of about \SI{150}{\nano\kelvin}, the average relative kinetic energy of the pairs translates into a small thermal frequency shift $(3/2)k_BT/h$ of about \SI{5}{\kilo\hertz}.
The strength and the duration of the modulation are adjusted in order to optimize the signal, the width and amplitude of which depend on the measurement. A Gaussian fit to the loss features yields widths (2 standard deviations) of 10 to \SI{20}{\kilo\hertz} and amplitudes ranging from 30 to 100\% of the atom number of the minority component (potassium).
In Fig.\,\ref{Fig4:Wiggle} we plot the binding energies measured as a function of the magnetic field. The data are fitted with Eq.\,(\ref{Eq:BindingEnergy}).
The fit results for the resonance parameters $B_0,\ \delta B^*$ and $\delta\mu$ are reported in Table \,\ref{Table:FBParameter} and discussed in Sec.\,\ref{SubSec:FBParametersDiscussion}.

 As for all the other measurements reported in this article, the magnetic field is calibrated performing RF spectroscopy on the atomic K transition $\ket{F,m_f}=\ket{\frac{9}{2},-\frac{9}{2}}\rightarrow\ket{\frac{9}{2},-\frac{7}{2}}$. The signal is fitted with a Lorentzian function and the uncertainty typically amounts to about \SI{1.5}{\milli\gauss}, as described in Ref.\,\cite{Ye2022OOL}.
 
\subsection{Magnetic moment spectroscopy}\label{SubSec:MagnMomentSpect}
\begin{figure}
\centering
\includegraphics[
width=1\columnwidth
]{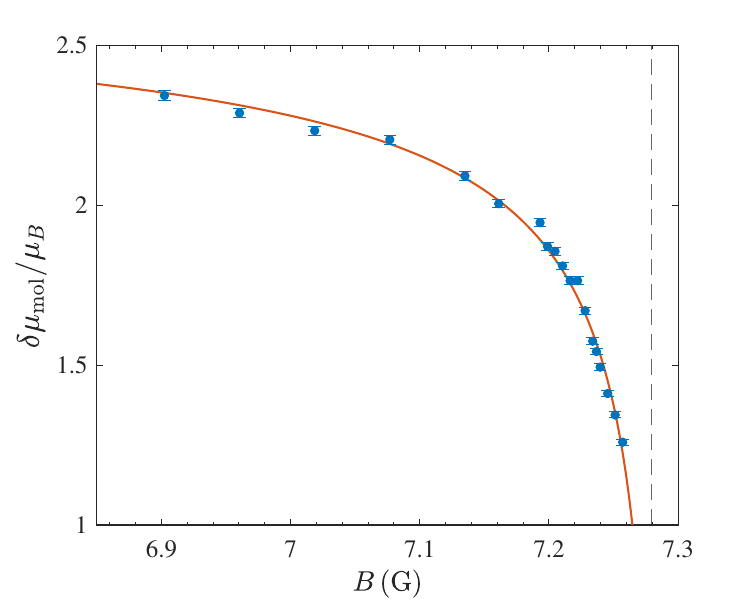}
\caption{Differential magnetic moment of the molecules as a function of the magnetic field. The magnetic moment is measured by determining the perfect magnetic field gradient that balances the gravitational force at the given magnetic field. The solid line is a fit to the data according to Eq.\,(\ref{Eq:MagneticMoment}). The dashed line indicates the position of the resonance pole.
}\label{Fig5:MagneticMoment}
\end{figure}
Applying an alternative method, we determine the resonance parameters by magnetic moment spectroscopy \cite{Herbig2003poa, Mark2007SOU}. 
The molecules are prepared as described in Sec.\,\ref{Sec:PreparationSection}. After the SG purification phase, we ramp the magnetic field to the target field. The magnetic field gradient is ramped accordingly to always allow the molecules to stay in the trap. At the target field we hold the molecules for \SI{4}{\milli\second} and then switch off the ODT. Both the magnetic field and the magnetic field gradient remain unchanged. We observe the vertical position of the molecular cloud during the expansion and determine the magnetic field gradient that perfectly compensates the gravitational force. The magnetic field gradient is calibrated performing corresponding measurements on Dy alone, the magnetic moment of which is well known ($\mu_{\mathrm{Dy}}=10\mu_B$, where $\mu_B$ is the Bohr magneton). 
In Fig.\,\ref{Fig5:MagneticMoment} we plot the differential molecular magnetic moment $\delta\mu_{\mathrm{mol}}$ as a function of the magnetic field strength, the magnetic moment $\mu_{\mathrm{K}}$ of potassium being $1\mu_B$. We fit the experimental data with Eq.\,(\ref{Eq:MagneticMoment}).
The parameter values derived from the fit are reported in Table \ref{Table:FBParameter}.

We point out that magnetic moment spectroscopy is a practical and powerful tool. It does not only provide us with an alternative way to determine the Feshbach resonance parameters, but also gives a direct measurement of the closed-channel fraction \cite{Chin2010FRI, Falco2005AMT}
\begin{equation}\label{Eq:ClosedChannelFraction}
Z(B)=\delta\mu_{\mathrm{mol}}(B)/ \delta\mu.
\end{equation}
We note that magnetic moment spectroscopy offers a great practical advantage. Knowing the resonance parameters $\delta B^*$ and $\delta\mu$, a single measurement of the magnetic gradient that balances gravity directly provides the magnetic detuning $\delta B$ and thus also the scattering length close to resonance.

\subsection{Summary of the 7-G resonance parameters}\label{SubSec:FBParametersDiscussion}
\begin{table}[b]
\begin{ruledtabular}
\caption{\label{Table:FBParameter} Summary of the parameter values determined for the 7-G Feshbach resonance.}
\begin{tabular}{lccc}
\textrm{}&
$\delta B^*\,$(\SI{}{\milli\gauss})&
$\delta\mu/\mu_B$ &
$B_0\,$(\SI{}{\gauss})\\
\colrule
\textrm{Binding energy spectroscopy} & 9.8(7) & 2.77(3) & 7.274(1) \\
\textrm{Magnetic moment spectroscopy} & 10.2(1.2) & 2.81(5) & 7.279(2) \\
Combined & 9.9(6) & 2.78(2) & 7.276(2) \\
\end{tabular}
\end{ruledtabular}
\end{table}
The values of the Feshbach resonance parameters $\delta B^*, \delta\mu$ and $B_0$ obtained from the binding energy spectroscopy and the magnetic moment spectroscopy are reported in Table \ref{Table:FBParameter}. The two methods give consistent results. We also present combined values for the three parameters, which represent our best knowledge.

The resonance center position derived in the present work is \SI{20}{\milli\gauss} lower than the value of $B_0=7.295(1)$ \SI{}{\gauss} reported in our previous work \cite{Ye2022OOL}.
We attribute the discrepancy between the two values to small uncontrolled magnetic field deviations affecting the previous measurements. The experiments reported in the present work have been conducted under improved magnetic field control. In particular, the combination of the fast coils and the digital proportional-integral-derivative (PID) controller leads to a faster stabilization of the magnetic field around the desired set point, minimizing uncontrolled departures of the magnetic field from the target value. Our present results for the parameter values of $\delta B^*$ and $\delta\mu$ are consistent with the previous results within one and two standard errors, respectively.


For the two parameters $R^*$ and $A$, which can be calculated from $\delta B^*$ and $\delta\mu$, we obtain the updated values $R^* = 604(20) a_0$ and $A = 24.0(6)$ G.


\section{Magnetic antitrapping in levitation field}\label{Sec:Antitrapping}
\begin{figure}
\centering
\includegraphics[
width=1\columnwidth
]{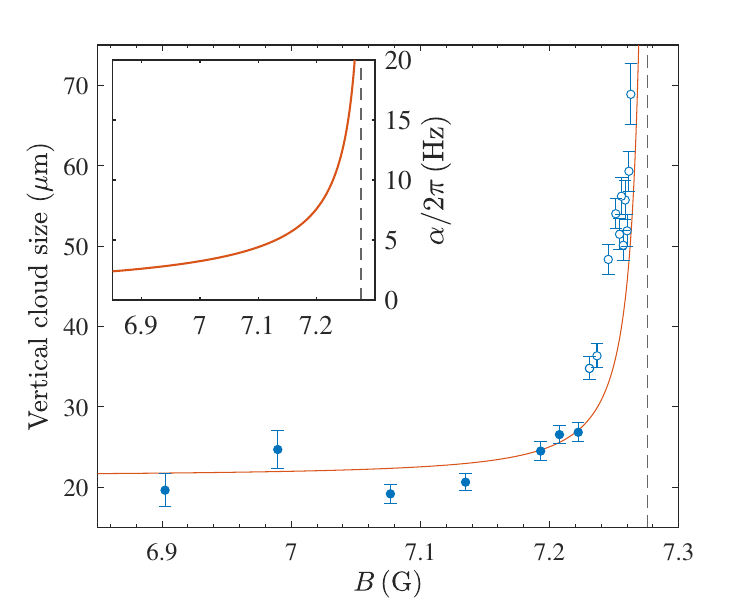}
\caption{
Measured vertical size of the levitated molecular cloud after {\SI{20}{\milli\second}} expansion as a function of the magnetic field. The error bars show the sample standard deviation of four individual measurements at the same magnetic field. The solid line is a fit to the data at the large detunings (filled circles) according to Eq.\,(\ref{Eq:VerticalDeformationTOF}).
The inset shows the antitrapping frequency $\alpha/2\pi$ calculated according to Eq.\,(\ref{Eq:AntitrapFreq}). For both plots the dashed line indicates the position of the pole of the resonance.}\label{Fig6:StretchingEffect}
\end{figure}
The magnetic field dependence of the molecular magnetic moment translates, in the presence of a magnetic field gradient, into a position-dependent force. For perfect levitation with a constant magnetic gradient along the vertical axis, this force balances gravity in the trap center, while it pushes up the molecules above the center and pulls down the ones below the center.
In first order of the vertical position $z$ with respect to the trap center, this anti-trapping force can be expressed as
\begin{equation}
    F(z) = (\partial_z \mu_{\rm mol})
    (\partial_z B) \, z \, ,
\end{equation}
where $\mu_{\rm mol} = \mu_{\rm Dy} + \mu_{\rm K} - \delta \mu_{\rm mol}$ is the total molecular magnetic moment.
With $\partial_z \mu_{\rm mol} =  (\partial_B \mu_{\rm mol}) (\partial_z B)$ and
$\partial_B \mu_{\rm mol} = -\partial^2_B E_{\rm mol}$ the force can be rewritten in terms of the second derivative of the binding energy:
\begin{equation}
    F(z) = -(\partial^2_z E_{\rm mol})
    (\partial_z B)^2 \, z \, .
\end{equation}
Interpreting the positive factor $-(\partial^2_z E_{\rm mol})
    (\partial_z B)^2 $ as a negative spring constant,
we now define the ``anti-trapping frequency''
\begin{equation}
    \alpha = \sqrt{-m_{\rm mol}^{-1} \, \partial^2_z E_{\rm mol}} \, \partial_z B \, ,
\end{equation}
where $m_{\rm mol} = m_{\rm Dy} + m_{\rm K}$ is the molecular mass.
Combining with Eq.\,\ref{Eq:BindingEnergy}) and applying some straightforward algebra, we finally obtain

\begin{equation}\label{Eq:AntitrapFreq}
\alpha=\sqrt{\frac{\delta\mu}{2m_{\rm mol}\delta B^*}}\frac{\partial_zB}{\left(1-\frac{\delta B}{\delta B^*}\right)^{3/4}}.
\end{equation}
In the resonance limit $\delta B \rightarrow 0$, where the levitation gradient takes the value $\partial_z B = m_{\rm mol} g / (\mu_{\rm Dy} + \mu_{\rm K}$) = 3.21\,G/cm, the anti-trapping frequency approaches its maximum $\alpha_{\rm max} = g \sqrt{\delta \mu\,m_{\rm mol} / (2 \delta B^*)} \, /(\mu_{\rm Dy}+\mu_{\rm K}) = 2\pi \times \SI{31.8(8)}{\hertz}$. The calculated anti-trapping frequency is shown in the inset of Fig.\,\ref{Fig6:StretchingEffect}: The closer to resonance, the stronger is the antitrapping force. At detunings larger than about \SI{50}{\milli\gauss}, $\alpha$ decreases to few \SI{}{\hertz}, with negligible effects both on the expansion dynamics and the trapping potential.

The magnetic anti-trapping effect can have considerable impact on molecules in the ODT. The vertical trap frequency is reduced from its bare value $\omega_z$ to an effective value $\sqrt{\omega^2_z - \alpha^2}$. Moreover, with the harmonic anti-trap competing with the Gaussian potential of the ODT, the overall depth of the trapping potential can be substantially reduced.
When the ODT is switched off, the antitrapping force modifies the ballistic expansion, causing a vertical stretching of the cloud. After release, the particle motion along the vertical direction is described in terms of hyperbolic functions \cite{Weber2003BEC,Herbig2003poa}. Assuming a Gaussian distribution of the position and velocity of the molecular sample, the vertical rms size of the cloud is expected to expand as
\begin{equation}\label{Eq:VerticalDeformationTOF}
\sigma_z=\sqrt{\frac{k_BT}{m_{\mathrm{mol}}}}\sqrt{\frac{\cosh^2(\alpha t_{\mathrm{TOF}})}{\omega_z^2-\alpha^2}+\frac{\sinh^2(\alpha t_{\mathrm{TOF}})}{\alpha^2}},
\end{equation}
where $T$ is the temperature of the sample and $t_{\mathrm{TOF}}$ the expansion time.

The stretching effect is shown in Fig.\,\ref{Fig6:StretchingEffect}, where the vertical rms size of the molecular cloud after $t_{\rm TOF}=$ \SI{20}{\milli\second} of free expansion is plotted as a function of the magnetic field. 
The closer to the resonance pole, the more pronounced is the cloud elongation. For a magnetic field detuning larger than about \SI{50}{\milli\gauss}, the effect becomes less relevant and $\sigma_z$ approaches the ballistic-expansion value of about \SI{20}{\micro\meter}. Restricting our attention to the large detuning region, $|\delta B|>$\,\SI{50}{\milli \gauss}, we fit the data based on Eq.\,(\ref{Eq:VerticalDeformationTOF}) with $T$ being the only free parameter. The resulting value of $T=\SI{27(5)}{\nano\kelvin}$ is somewhat lower than what was obtained from other time of flight measurements. We attribute this discrepancy to a measurement artefact: The cloud was held in the ODT for a time interval that accidentally coincided with a minimum velocity spread during a breathing mode oscillation.
Overall, the observed vertical stretching effect follows Eq.\,(\ref{Eq:VerticalDeformationTOF}) well, which confirms our model of an anti-trapping effect described by the frequency $\alpha$ according to Eq.\,(\ref{Eq:AntitrapFreq}); see also inset in Fig.\,\ref{Fig6:StretchingEffect}. Close to resonance the anti-trapping effect can substantially modify the free expansion. Only at larger magnetic field detuning ($|\delta B|\gg \delta B^*$) the effect becomes negligible.

\section{Lifetime}\label{Sec:Lifetime}
\begin{figure*}
\centering
\includegraphics[
width=2\columnwidth
]{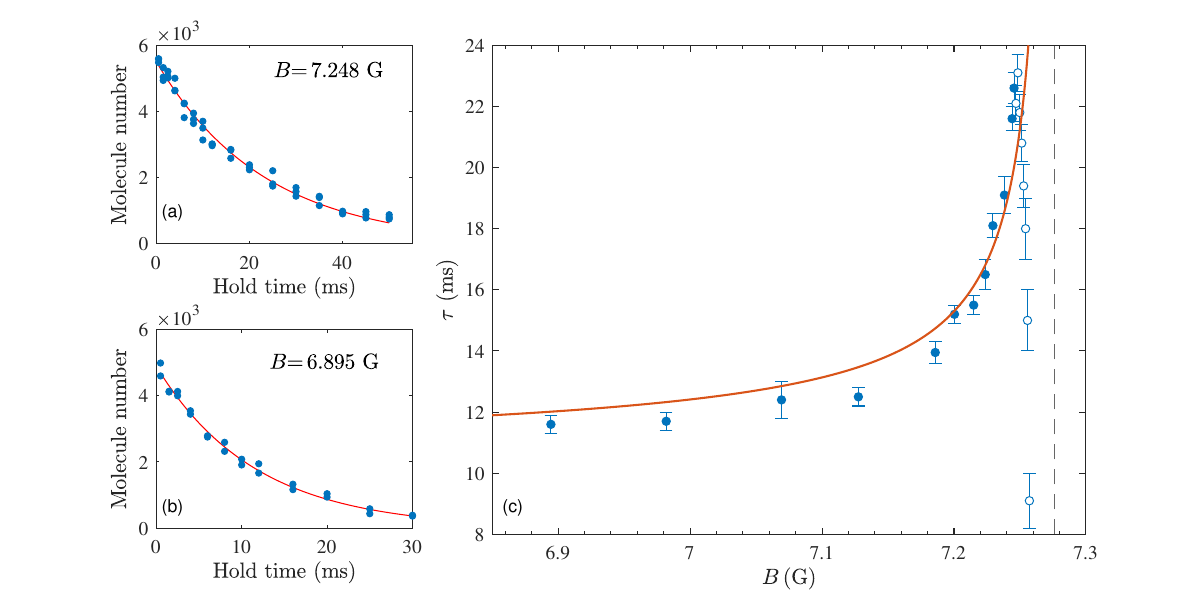}
\caption{Decay of the pure trapped molecular sample. Typical decay curves are shown in panel (a) and (b); note the different time scales. The lifetimes have been measured in an ODT with trap frequency $\bar{\omega}\approx2\pi\times\SI{20}{\hertz}$, at a magnetic detuning of \SI{-27}{\milli\gauss} (a) and \SI{-380}{\milli\gauss} (b). From an exponential fit to the data (solid lines) we derive decay times of \SI{23.1(6)}{} and \SI{11.6(3)}{\milli\second}, respectively. In panel (c) we report decay times measured under the same experimental conditions, as function of the magnetic field strength. The solid line represents the result of a one-parameter fit (see text) to the data at large detunings ($|\delta B|>\SI{30}{\milli\gauss}$, filled circles). The dashed line indicates the position of the pole of the resonance.
}\label{Fig7:Lifetime} 
\end{figure*}
The stability of the molecular cloud against losses is of primary importance for many future experiments with our system. The fact that we work with a pure sample of molecules already eliminates collisions with free atoms as possible sources of losses. Inelastic collisions between the dimers and processes induced by the trap light thus remain as possible sources. In this section, we study the decay of the optically trapped dimer sample. We measure lifetimes up to 20\,ms, and we identify trap-light induced losses as the dominant decay mechanism.

The molecular sample is prepared as described in Sec.\,\ref{Sec:PreparationSection}. At the end of the purification phase, the cloud is transferred to the target field and the magnetic gradient is ramped accordingly. The molecules are held in the ODT for a variable time, after which they are released and let expand in time of flight. The molecules are dissociated \SI{1}{\milli\second} before the imaging pulse. 
Typical decay curves can be observed in Fig.\,\ref{Fig7:Lifetime}(a) and (b), measured in a rather shallow trap ($\bar{\omega}\approx 2\pi\times\SI{20}{\hertz}$) at $B=\SI{7.248}{}$ and \SI{6.895}{\gauss}, respectively. 
The decay follows a simple exponential behavior, suggesting that the limiting loss process is a one-body mechanism. As the only plausible explanation, we assume that the infrared light of the ODT induces transitions into electronically excited molecular states \cite{Partridge2005MPO}. 

The magnetic field detuning clearly has an impact on the lifetime of the sample: For the smaller detuning we measure a decay time almost two times longer than for the larger detuning.
We systematically investigate the dependence of the lifetime on the magnetic field. The decay times $\tau$ reported in Fig.\,\ref{Fig7:Lifetime}(c) are derived from exponential fits to the corresponding lifetime curves. All curves have been recorded under essentially the same trapping conditions as for Figs.\,\ref{Fig7:Lifetime}(a) and (b), with an initial molecular number varying between 5 and \SI{6e3}{}.
At large detunings the sample features a relatively fast decay, with $\tau\approx\SI{12}{\milli\second}$. The lifetime of the sample rapidly increases when approaching the pole of the resonance, reaching decay times up to $\tau=\SI{23}{\milli\second}$ for a magnetic field detuning of about \SI{30}{\milli\gauss}.

To explain the behavior we assume that the closed-channel fraction of the Feshbach molecules is responsible for the losses \cite{Partridge2005MPO}. The optical excitation presumably couples the closed channel to some unknown electronically excited molecular states.
We therefore model the lifetime by
\begin{equation}\label{Eq:TauMolecules}
\tau(B) = \tau_{\rm closed}/Z(B),
\end{equation}
where $Z$ is the closed-channel fraction according to Eq.\,(\ref{Eq:ClosedChannelFraction}) and $\tau_{\rm closed}$ is the lifetime of the closed-channel molecule. In Fig.\,\ref{Fig7:Lifetime}(c) we show the corresponding behavior of $\tau(B)$ calculated with the resonance parameters as determined before and obtaining $\tau_{\rm closed} = \SI{10.1 (4)}{\milli\second}$ from a single free parameter fit to the data for     $|\delta B|>\SI{30}{\milli\gauss}$. 


For $|\delta B|\lesssim \SI{30}{\milli\gauss}$, the observed lifetime changes dramatically, with $\tau$ rapidly decreasing while approaching the resonance pole.
This different behavior can be understood in terms of the anti-trapping mechanism introduced in Sec.\,\ref{Sec:Antitrapping}, which, close to to resonance pole, becomes sufficiently strong to effectively pull the molecules out of the weak ODT, thus reducing the lifetime of the sample.

\begin{figure}
\centering
\includegraphics[
width=1\columnwidth
]{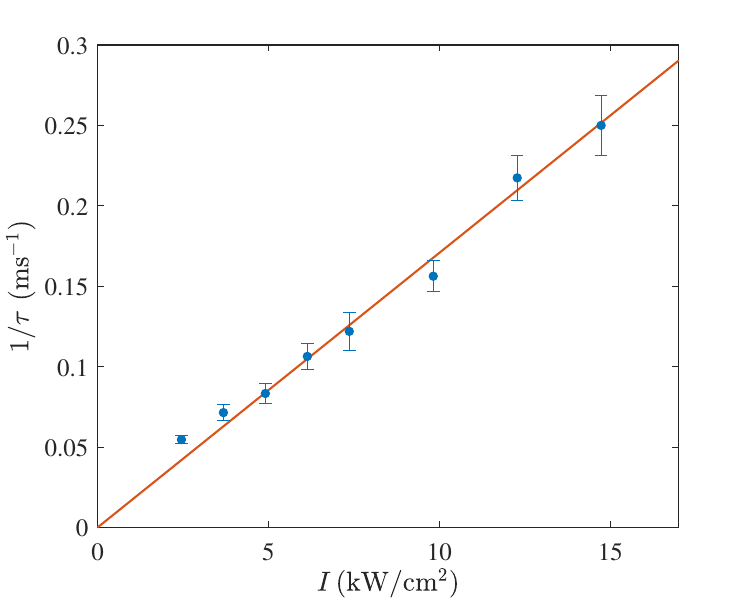}
\caption{Linear dependence of the measured loss rate on the beam intensity. The rates have been measured in a single-beam ODT with a waist of \SI{90}{\micro\meter} and a wavelength of $\SI{1064}{\nano\meter}$. All data have been taken at a constant magnetic field detuning of about \SI{30}{\milli\gauss}. A linear fit to the data (solid line) yields a slope of \SI{1.71(3)e-2}{\centi\meter^2\second^{-1}/\kilo\watt}.
}\label{Fig8:LifetimeVsIntensity} 
\end{figure}

To further verify that the observed losses are induced by the trap light, we measure the dependence of the loss rate on the intensity \textit{I} of the trapping beam. The molecules are held in a single-beam ODT in the presence of a levitating magnetic field gradient. The lifetime curves have been recorded for different trap intensities, at a fixed magnetic field $B=\SI{7.244}{\gauss}$. Even at high intensity, for which the trap becomes very tight, the decay curves preserve a purely exponential behavior, suggesting that in the region of system parameters explored no density-dependent effects enter into play.    
Our results are reported in Fig.\,\ref{Fig8:LifetimeVsIntensity} and show a linear dependence of the loss rate on the beam intensity. This linearity indicates that the loss mechanism is not saturated, which suggests an off-resonant process to take place instead of a resonant excitation of a specific transition.

Following basically the same protocols, we have also carried out measurements on the lifetime of trapped Feshbach molecules produced via the resonance near \SI{51.2}{\gauss}, which has a very similar character to the 7.3-G resonance \cite{Ye2022OOL}. Here we observed a much faster light-induced decay on the timescale of one millisecond. This shows that the decay depends very sensitively on the particular closed-channel molecular state underlying the Feshbach resonance, and on its coupling to electronically excited states.

Light-induced one-body losses of optically trapped Feshbach molecules have been observed in Ref.\,\cite{Chotia2012lld}. Here single KRb molecule were trapped in individual sites of an optical lattice, which shielded them and thus suppressed collisional losses. The dependence of decay times on the closed-channel fraction was also studied with similar findings as in our experiment. Trap-light-induced losses were also investigated as possible limitations to experiments with single Feshbach molecules of NaCs or RbCs in optical tweezers \cite{Zhang2020FSM, Spence2023ASR}. Decay measurements on samples of optically trapped pure Feshbach molecules of RbCs \cite{Koeppinger2014poo}, very similar to our experiments, also showed an exponential loss behavior, which we interpret as a result of dominant one-body losses. Such trap-light induced losses seem to be rather the rule than the exception, although they have not received much attention so far.


\section{Conclusions and outlook}
\label{Sec:Outro}

We have demonstrated the preparation of a pure sample of ultracold DyK Feshbach molecules in an optical dipole trap, with high-phase space densities of the order of 0.1. The weakly bound molecules are composed of the fermionic isotopes $^{161}$Dy and $^{40}$K, which makes the sample an interesting starting point for experiments addressing the few- and many-body physics of resonantly interacting fermions with mass imbalance. 

The efficient purification scheme applied in our experiments is based on a Stern-Gerlach method, where magnetic levitation keeps the molecules in a shallow trap, while the atoms are removed by the combined action of gravity and the magnetic gradient force. We have shown that the measured dependence of the molecular magnetic moment on the magnetic field strength provides an accurate characterization of the underlying Feshbach resonance and complements spectroscopic measurements of the molecular binding energy. We have also demonstrated a peculiar effect appearing in the free expansion of the molecular cloud in the magnetic gradient field. The sharp dependence of the magnetic moment in the universal regime close to the resonance center manifests itself in a vertical antitrapping force, which leads to an anisotropy (vertical stretching) of the expanding cloud.

Lifetime studies have revealed a one-body mechanism induced by the trap light as the main limitation. We have observed that the decay rate is proportional to the closed-channel fraction of the Feshbach molecules and also proportional to the trap light intensity. This suggests an off-resonant photoexcitation mechanism that couples the closed-channel molecular state to the manifold of electronically excited states and leads to losses, possibly by photodissociation or optical pumping. Understanding and suppressing this effect needs more investigations, in particular by variation of the trap wavelength. Tuning the trap light, which was not feasible in our present setup employing fixed-frequency laser sources, will allow us to mitigate and possibly cancel photo-excitation loss processes. Generally, it can be expected that using trap light much further in the infrared than the standard 1064-nm wavelength (e.g.\ in the \SI{1550}{\nano\meter} or \SI{2}{\micro\meter} regions) will strongly reduce these losses because of a decreasing density of electronically excited molecular states that can couple to the Feshbach dimer.

Reaching quantum degeneracy with the molecular sample and thus the observation of a BEC of heteronuclear Feshbach molecules will presumably require a final stage of evaporative cooling. Under the current conditions, however, it can be estimated the a molecule only undergoes roughly one elastic collision in the limited lifetime of the sample. Overcoming this limitation demands both a substantial improvement of the lifetimes (by optimizing the trap wavelength) and an increase in the elastic collision rate. The latter can be achieved by working at small magnetic detunings very close to the resonance center. Deep in the universal regime, the molecular $s$-wave scattering length is predicted to follow the atomic scattering length, $a_{\rm mol} \approx 0.77\,a$ \cite{Petrov2005dmi}, and will thus also be resonantly enhanced. Moreover, the increasing Pauli blocking of inelastic collisions near resonance \cite{Petrov2004wbd, Petrov2005dmi} will strongly suppress collisional losses, which is a well-established effect in spin-mixtures of fermions \cite{Inguscio2008ufg} and has also been observed in a heteronuclear K-Li Fermi-Fermi mixture \cite{Jag2016lof}. The magnetic antitrapping effect that is caused by the levitation field may be overcome by, e.g., using a standing-wave trap that provides tight vertical confinement. Taking advantage of all these potential improvements, we are confident that a sufficiently high ratio of elastic vs. inelastic molecule-molecule collisions can be reached to implement efficient evaporative cooling towards quantum degeneracy.

\begin{acknowledgments}
 
We thank M.\ Zaccanti for stimulating discussions and for comments on the manuscript. We also thank D.\ Petrov and S.\ Cornish for insightful discussions and Y.\ Yudkin for contributions
in the early stage of this work. The project has received funding from the European Research Council (ERC) under the European Union’s Horizon 2020 research and innovation programme (grant agreement No.\ 101020438 - SuperCoolMix). We further acknowledge support by the Austrian Science Fund (FWF) within the Doktoratskolleg ALM (W1259-N27).  We thank the members of the ultracold atom groups in Innsbruck for many stimulating discussions and for sharing technological know-how.
 
\end{acknowledgments}
~\\

\appendix
\section{Magnetic field control}\label{App:MagneticFieldControl}
Precise magnetic field control is crucial for an efficient production of the molecular sample, as well as for the precise determination of the Feshbach resonance parameters. In this regard, we put a lot of efforts into the implementation of an effective magnetic-field ramping protocol, as well as into the development of a precise current actuator for the coils that produce the magnetic field. The Feshbach magnetic field is produced employing two pairs of coils. 

The offset magnetic field $B_{\rm{off}}$ is generated by a slow set of coils ({offset coils}), whose proportional-integral-derivative (PID) actuator settings are optimized to perform the large magnetic field ramp within \SI{2}{\milli\second}. 
Every magnetic field ramp unavoidably induces some eddy currents in the apparatus, which delay the response of the actual magnetic field with respect to the targeted behavior. We minimize this undesired effect by applying a feed-forward scheme on the current ramp. Furthermore, we wait about 50 ms after the ramp, to allow the magnetic field to fully stabilize to the set point. The magnetic field noise is measured to be below \SI{2}{\milli\gauss} peak to peak.
The residual magnetic-field produced by the offset coils provides a fixed stable offset, which is left unvaried for the rest of the sequence.

The magnetic field ramps discussed in Sec.\,\ref{SubSec:SGPurification} are performed by means of a low-inductance coils ({fast coils}) whose magnetic field is superimposed with $B_{\rm{off}}$ provided by the offset coils. The fast coils can handle a maximum current of about \SI{10}{\ampere}, corresponding to about \SI{3}{\gauss}. The current flow is regulated with a metal–oxide–semiconductor field-effect transistor (MOSFET) whose gate voltage is controlled digitally via a field programmable gate array (FPGA). Following the work reported in Ref. \cite{Thomas2020ADF}, the FPGA is programmed to optimize the feedback loop and to set the PID parameters accordingly during the current ramp. The quality of the digital PID controller and the low current-to-field conversion factor are such that the {fast coils} do not introduce extra magnetic field noise in the system.

\bibliography{ultracold, BibliographyProductionPureMolecules}

\begin{thebibliography}{73}%
\makeatletter
\providecommand \@ifxundefined [1]{%
 \@ifx{#1\undefined}
}%
\providecommand \@ifnum [1]{%
 \ifnum #1\expandafter \@firstoftwo
 \else \expandafter \@secondoftwo
 \fi
}%
\providecommand \@ifx [1]{%
 \ifx #1\expandafter \@firstoftwo
 \else \expandafter \@secondoftwo
 \fi
}%
\providecommand \natexlab [1]{#1}%
\providecommand \enquote  [1]{``#1''}%
\providecommand \bibnamefont  [1]{#1}%
\providecommand \bibfnamefont [1]{#1}%
\providecommand \citenamefont [1]{#1}%
\providecommand \href@noop [0]{\@secondoftwo}%
\providecommand \href [0]{\begingroup \@sanitize@url \@href}%
\providecommand \@href[1]{\@@startlink{#1}\@@href}%
\providecommand \@@href[1]{\endgroup#1\@@endlink}%
\providecommand \@sanitize@url [0]{\catcode `\\12\catcode `\$12\catcode
  `\&12\catcode `\#12\catcode `\^12\catcode `\_12\catcode `\%12\relax}%
\providecommand \@@startlink[1]{}%
\providecommand \@@endlink[0]{}%
\providecommand \url  [0]{\begingroup\@sanitize@url \@url }%
\providecommand \@url [1]{\endgroup\@href {#1}{\urlprefix }}%
\providecommand \urlprefix  [0]{URL }%
\providecommand \Eprint [0]{\href }%
\providecommand \doibase [0]{https://doi.org/}%
\providecommand \selectlanguage [0]{\@gobble}%
\providecommand \bibinfo  [0]{\@secondoftwo}%
\providecommand \bibfield  [0]{\@secondoftwo}%
\providecommand \translation [1]{[#1]}%
\providecommand \BibitemOpen [0]{}%
\providecommand \bibitemStop [0]{}%
\providecommand \bibitemNoStop [0]{.\EOS\space}%
\providecommand \EOS [0]{\spacefactor3000\relax}%
\providecommand \BibitemShut  [1]{\csname bibitem#1\endcsname}%
\let\auto@bib@innerbib\@empty
\bibitem [{\citenamefont {K\"ohler}\ \emph {et~al.}(2006)\citenamefont
  {K\"ohler}, \citenamefont {Goral},\ and\ \citenamefont
  {Julienne}}]{Koehler2006poc}%
  \BibitemOpen
  \bibfield  {author} {\bibinfo {author} {\bibfnamefont {T.}~\bibnamefont
  {K\"ohler}}, \bibinfo {author} {\bibfnamefont {K.}~\bibnamefont {Goral}},\
  and\ \bibinfo {author} {\bibfnamefont {P.~S.}\ \bibnamefont {Julienne}},\
  }\bibfield  {title} {\bibinfo {title} {Production of cold molecules via
  magnetically tunable {F}eshbach resonances},\ }\href
  {https://doi.org/10.1103/RevModPhys.78.1311} {\bibfield  {journal} {\bibinfo
  {journal} {Rev. Mod. Phys.}\ }\textbf {\bibinfo {volume} {78}},\ \bibinfo
  {pages} {1311} (\bibinfo {year} {2006})}\BibitemShut {NoStop}%
\bibitem [{\citenamefont {Chin}\ \emph {et~al.}(2010)\citenamefont {Chin},
  \citenamefont {Grimm}, \citenamefont {Julienne},\ and\ \citenamefont
  {Tiesinga}}]{Chin2010FRI}%
  \BibitemOpen
  \bibfield  {author} {\bibinfo {author} {\bibfnamefont {C.}~\bibnamefont
  {Chin}}, \bibinfo {author} {\bibfnamefont {R.}~\bibnamefont {Grimm}},
  \bibinfo {author} {\bibfnamefont {P.~S.}\ \bibnamefont {Julienne}},\ and\
  \bibinfo {author} {\bibfnamefont {E.}~\bibnamefont {Tiesinga}},\ }\bibfield
  {title} {\bibinfo {title} {Feshbach resonances in ultracold gases},\ }\href
  {https://doi.org/doi.org/10.1103/RevModPhys.82.1225} {\bibfield  {journal}
  {\bibinfo  {journal} {Rev. Mod. Phys.}\ }\textbf {\bibinfo {volume} {82}},\
  \bibinfo {pages} {1225} (\bibinfo {year} {2010})}\BibitemShut {NoStop}%
\bibitem [{\citenamefont {Wynar}\ \emph {et~al.}(2000)\citenamefont {Wynar},
  \citenamefont {Freeland}, \citenamefont {Han}, \citenamefont {Ryu},\ and\
  \citenamefont {Heinzen}}]{Wynar2000mia}%
  \BibitemOpen
  \bibfield  {author} {\bibinfo {author} {\bibfnamefont {R.}~\bibnamefont
  {Wynar}}, \bibinfo {author} {\bibfnamefont {R.~S.}\ \bibnamefont {Freeland}},
  \bibinfo {author} {\bibfnamefont {D.~J.}\ \bibnamefont {Han}}, \bibinfo
  {author} {\bibfnamefont {C.}~\bibnamefont {Ryu}},\ and\ \bibinfo {author}
  {\bibfnamefont {D.~J.}\ \bibnamefont {Heinzen}},\ }\bibfield  {title}
  {\bibinfo {title} {{Molecules in a Bose-Einstein Condensate}},\ }\href
  {https://doi.org/10.1126/science.287.5455.1016} {\bibfield  {journal}
  {\bibinfo  {journal} {Science}\ }\textbf {\bibinfo {volume} {287}},\ \bibinfo
  {pages} {1016} (\bibinfo {year} {2000})}\BibitemShut {NoStop}%
\bibitem [{\citenamefont {Donley}\ \emph {et~al.}(2002)\citenamefont {Donley},
  \citenamefont {Clausen}, \citenamefont {Thompson},\ and\ \citenamefont
  {Wieman}}]{Donley2002amc}%
  \BibitemOpen
  \bibfield  {author} {\bibinfo {author} {\bibfnamefont {E.~A.}\ \bibnamefont
  {Donley}}, \bibinfo {author} {\bibfnamefont {N.~R.}\ \bibnamefont {Clausen}},
  \bibinfo {author} {\bibfnamefont {S.~T.}\ \bibnamefont {Thompson}},\ and\
  \bibinfo {author} {\bibfnamefont {C.~E.}\ \bibnamefont {Wieman}},\ }\bibfield
   {title} {\bibinfo {title} {{Atom-molecule coherence in a Bose-Einstein
  condensate}},\ }\href {https://doi.org/10.1038/417529a} {\bibfield  {journal}
  {\bibinfo  {journal} {Nature (London)}\ }\textbf {\bibinfo {volume} {417}},\
  \bibinfo {pages} {529} (\bibinfo {year} {2002})}\BibitemShut {NoStop}%
\bibitem [{\citenamefont {Herbig}\ \emph {et~al.}(2003)\citenamefont {Herbig},
  \citenamefont {Kraemer}, \citenamefont {Mark}, \citenamefont {Weber},
  \citenamefont {Chin}, \citenamefont {N\"agerl},\ and\ \citenamefont
  {Grimm}}]{Herbig2003poa}%
  \BibitemOpen
  \bibfield  {author} {\bibinfo {author} {\bibfnamefont {J.}~\bibnamefont
  {Herbig}}, \bibinfo {author} {\bibfnamefont {T.}~\bibnamefont {Kraemer}},
  \bibinfo {author} {\bibfnamefont {M.}~\bibnamefont {Mark}}, \bibinfo {author}
  {\bibfnamefont {T.}~\bibnamefont {Weber}}, \bibinfo {author} {\bibfnamefont
  {C.}~\bibnamefont {Chin}}, \bibinfo {author} {\bibfnamefont {H.-C.}\
  \bibnamefont {N\"agerl}},\ and\ \bibinfo {author} {\bibfnamefont
  {R.}~\bibnamefont {Grimm}},\ }\bibfield  {title} {\bibinfo {title}
  {Preparation of a pure molecular quantum gas},\ }\href
  {https://doi.org/10.1126/science.1088876} {\bibfield  {journal} {\bibinfo
  {journal} {Science}\ }\textbf {\bibinfo {volume} {301}},\ \bibinfo {pages}
  {1510} (\bibinfo {year} {2003})}\BibitemShut {NoStop}%
\bibitem [{\citenamefont {Xu}\ \emph {et~al.}(2003)\citenamefont {Xu},
  \citenamefont {Mukaiyama}, \citenamefont {Abo-Shaeer}, \citenamefont {Chin},
  \citenamefont {Miller},\ and\ \citenamefont {Ketterle}}]{Xu2003foq}%
  \BibitemOpen
  \bibfield  {author} {\bibinfo {author} {\bibfnamefont {K.}~\bibnamefont
  {Xu}}, \bibinfo {author} {\bibfnamefont {T.}~\bibnamefont {Mukaiyama}},
  \bibinfo {author} {\bibfnamefont {J.~R.}\ \bibnamefont {Abo-Shaeer}},
  \bibinfo {author} {\bibfnamefont {J.~K.}\ \bibnamefont {Chin}}, \bibinfo
  {author} {\bibfnamefont {D.~E.}\ \bibnamefont {Miller}},\ and\ \bibinfo
  {author} {\bibfnamefont {W.}~\bibnamefont {Ketterle}},\ }\bibfield  {title}
  {\bibinfo {title} {{Formation of Quantum-Degenerate Sodium Molecules}},\
  }\href {https://doi.org/10.1103/PhysRevLett.91.210402} {\bibfield  {journal}
  {\bibinfo  {journal} {Phys. Rev. Lett.}\ }\textbf {\bibinfo {volume} {91}},\
  \bibinfo {eid} {210402} (\bibinfo {year} {2003})}\BibitemShut {NoStop}%
\bibitem [{\citenamefont {D{\"u}rr}\ \emph
  {et~al.}(2004{\natexlab{a}})\citenamefont {D{\"u}rr}, \citenamefont {Volz},
  \citenamefont {Marte},\ and\ \citenamefont {Rempe}}]{Duerr2004oom}%
  \BibitemOpen
  \bibfield  {author} {\bibinfo {author} {\bibfnamefont {S.}~\bibnamefont
  {D{\"u}rr}}, \bibinfo {author} {\bibfnamefont {T.}~\bibnamefont {Volz}},
  \bibinfo {author} {\bibfnamefont {A.}~\bibnamefont {Marte}},\ and\ \bibinfo
  {author} {\bibfnamefont {G.}~\bibnamefont {Rempe}},\ }\bibfield  {title}
  {\bibinfo {title} {{Observation of Molecules Produced from a Bose-Einstein
  Condensate}},\ }\href {https://doi.org/10.1103/PhysRevLett.92.020406}
  {\bibfield  {journal} {\bibinfo  {journal} {Phys. Rev. Lett.}\ }\textbf
  {\bibinfo {volume} {92}},\ \bibinfo {eid} {020406} (\bibinfo {year}
  {2004}{\natexlab{a}})}\BibitemShut {NoStop}%
\bibitem [{\citenamefont {Regal}\ \emph {et~al.}(2003)\citenamefont {Regal},
  \citenamefont {Ticknor}, \citenamefont {Bohn},\ and\ \citenamefont
  {Jin}}]{Regal2003cum}%
  \BibitemOpen
  \bibfield  {author} {\bibinfo {author} {\bibfnamefont {C.~A.}\ \bibnamefont
  {Regal}}, \bibinfo {author} {\bibfnamefont {C.}~\bibnamefont {Ticknor}},
  \bibinfo {author} {\bibfnamefont {J.~L.}\ \bibnamefont {Bohn}},\ and\
  \bibinfo {author} {\bibfnamefont {D.~S.}\ \bibnamefont {Jin}},\ }\bibfield
  {title} {\bibinfo {title} {{Creation of ultracold molecules from a Fermi gas
  of atoms}},\ }\href {https://doi.org/10.1038/nature01738} {\bibfield
  {journal} {\bibinfo  {journal} {Nature (London)}\ }\textbf {\bibinfo {volume}
  {424}},\ \bibinfo {pages} {47} (\bibinfo {year} {2003})}\BibitemShut
  {NoStop}%
\bibitem [{\citenamefont {Strecker}\ \emph {et~al.}(2003)\citenamefont
  {Strecker}, \citenamefont {Partridge},\ and\ \citenamefont
  {Hulet}}]{Strecker2003coa}%
  \BibitemOpen
  \bibfield  {author} {\bibinfo {author} {\bibfnamefont {K.~E.}\ \bibnamefont
  {Strecker}}, \bibinfo {author} {\bibfnamefont {G.~B.}\ \bibnamefont
  {Partridge}},\ and\ \bibinfo {author} {\bibfnamefont {R.~G.}\ \bibnamefont
  {Hulet}},\ }\bibfield  {title} {\bibinfo {title} {{Conversion of an Atomic
  Fermi Gas to a Long-Lived Molecular Bose Gas}},\ }\href
  {https://doi.org/10.1103/PhysRevLett.91.080406} {\bibfield  {journal}
  {\bibinfo  {journal} {Phys. Rev. Lett.}\ }\textbf {\bibinfo {volume} {91}},\
  \bibinfo {pages} {080406} (\bibinfo {year} {2003})}\BibitemShut {NoStop}%
\bibitem [{\citenamefont {Cubizolles}\ \emph {et~al.}(2003)\citenamefont
  {Cubizolles}, \citenamefont {Bourdel}, \citenamefont {Kokkelmans},
  \citenamefont {Shlyapnikov},\ and\ \citenamefont
  {Salomon}}]{Cubizolles2003pol}%
  \BibitemOpen
  \bibfield  {author} {\bibinfo {author} {\bibfnamefont {J.}~\bibnamefont
  {Cubizolles}}, \bibinfo {author} {\bibfnamefont {T.}~\bibnamefont {Bourdel}},
  \bibinfo {author} {\bibfnamefont {S.~J. J. M.~F.}\ \bibnamefont
  {Kokkelmans}}, \bibinfo {author} {\bibfnamefont {G.~V.}\ \bibnamefont
  {Shlyapnikov}},\ and\ \bibinfo {author} {\bibfnamefont {C.}~\bibnamefont
  {Salomon}},\ }\bibfield  {title} {\bibinfo {title} {{Production of Long-Lived
  Ultracold Li$_2$ Molecules from a Fermi Gas}},\ }\href
  {https://doi.org/10.1103/PhysRevLett.91.240401} {\bibfield  {journal}
  {\bibinfo  {journal} {Phys. Rev. Lett.}\ }\textbf {\bibinfo {volume} {91}},\
  \bibinfo {pages} {240401} (\bibinfo {year} {2003})}\BibitemShut {NoStop}%
\bibitem [{\citenamefont {Jochim}\ \emph
  {et~al.}(2003{\natexlab{a}})\citenamefont {Jochim}, \citenamefont
  {Bartenstein}, \citenamefont {Altmeyer}, \citenamefont {Hendl}, \citenamefont
  {Chin}, \citenamefont {{Hecker Denschlag}},\ and\ \citenamefont
  {Grimm}}]{Jochim2003pgo}%
  \BibitemOpen
  \bibfield  {author} {\bibinfo {author} {\bibfnamefont {S.}~\bibnamefont
  {Jochim}}, \bibinfo {author} {\bibfnamefont {M.}~\bibnamefont {Bartenstein}},
  \bibinfo {author} {\bibfnamefont {A.}~\bibnamefont {Altmeyer}}, \bibinfo
  {author} {\bibfnamefont {G.}~\bibnamefont {Hendl}}, \bibinfo {author}
  {\bibfnamefont {C.}~\bibnamefont {Chin}}, \bibinfo {author} {\bibfnamefont
  {J.}~\bibnamefont {{Hecker Denschlag}}},\ and\ \bibinfo {author}
  {\bibfnamefont {R.}~\bibnamefont {Grimm}},\ }\bibfield  {title} {\bibinfo
  {title} {{Pure Gas of Optically Trapped Molecules Created from Fermionic
  Atoms}},\ }\href {https://doi.org/10.1103/PhysRevLett.91.240402} {\bibfield
  {journal} {\bibinfo  {journal} {Phys. Rev. Lett}\ }\textbf {\bibinfo {volume}
  {91}},\ \bibinfo {pages} {240402} (\bibinfo {year}
  {2003}{\natexlab{a}})}\BibitemShut {NoStop}%
\bibitem [{\citenamefont {Jochim}\ \emph
  {et~al.}(2003{\natexlab{b}})\citenamefont {Jochim}, \citenamefont
  {Bartenstein}, \citenamefont {Altmeyer}, \citenamefont {Hendl}, \citenamefont
  {Riedl}, \citenamefont {Chin}, \citenamefont {{Hecker Denschlag}},\ and\
  \citenamefont {Grimm}}]{Jochim2003bec}%
  \BibitemOpen
  \bibfield  {author} {\bibinfo {author} {\bibfnamefont {S.}~\bibnamefont
  {Jochim}}, \bibinfo {author} {\bibfnamefont {M.}~\bibnamefont {Bartenstein}},
  \bibinfo {author} {\bibfnamefont {A.}~\bibnamefont {Altmeyer}}, \bibinfo
  {author} {\bibfnamefont {G.}~\bibnamefont {Hendl}}, \bibinfo {author}
  {\bibfnamefont {S.}~\bibnamefont {Riedl}}, \bibinfo {author} {\bibfnamefont
  {C.}~\bibnamefont {Chin}}, \bibinfo {author} {\bibfnamefont {J.}~\bibnamefont
  {{Hecker Denschlag}}},\ and\ \bibinfo {author} {\bibfnamefont
  {R.}~\bibnamefont {Grimm}},\ }\bibfield  {title} {\bibinfo {title}
  {{Bose-Einstein Condensation of Molecules}},\ }\href
  {https://doi.org/10.1126/science.1093280} {\bibfield  {journal} {\bibinfo
  {journal} {Science}\ }\textbf {\bibinfo {volume} {302}},\ \bibinfo {pages}
  {2101} (\bibinfo {year} {2003}{\natexlab{b}})}\BibitemShut {NoStop}%
\bibitem [{\citenamefont {Greiner}\ \emph {et~al.}(2003)\citenamefont
  {Greiner}, \citenamefont {Regal},\ and\ \citenamefont
  {Jin}}]{Greiner2003eoa}%
  \BibitemOpen
  \bibfield  {author} {\bibinfo {author} {\bibfnamefont {M.}~\bibnamefont
  {Greiner}}, \bibinfo {author} {\bibfnamefont {C.~A.}\ \bibnamefont {Regal}},\
  and\ \bibinfo {author} {\bibfnamefont {D.~S.}\ \bibnamefont {Jin}},\
  }\bibfield  {title} {\bibinfo {title} {{Emergence of a molecular
  Bose-Einstein Condensate from a Fermi gas}},\ }\href
  {https://doi.org/doi.org/10.1038/nature02199} {\bibfield  {journal} {\bibinfo
   {journal} {Nature (London)}\ }\textbf {\bibinfo {volume} {426}},\ \bibinfo
  {pages} {537} (\bibinfo {year} {2003})}\BibitemShut {NoStop}%
\bibitem [{\citenamefont {Zwierlein}\ \emph {et~al.}(2003)\citenamefont
  {Zwierlein}, \citenamefont {Stan}, \citenamefont {Schunck}, \citenamefont
  {Raupach}, \citenamefont {Gupta}, \citenamefont {Hadzibabic},\ and\
  \citenamefont {Ketterle}}]{Zwierlein2003oob}%
  \BibitemOpen
  \bibfield  {author} {\bibinfo {author} {\bibfnamefont {M.~W.}\ \bibnamefont
  {Zwierlein}}, \bibinfo {author} {\bibfnamefont {C.~A.}\ \bibnamefont {Stan}},
  \bibinfo {author} {\bibfnamefont {C.~H.}\ \bibnamefont {Schunck}}, \bibinfo
  {author} {\bibfnamefont {S.~M.~F.}\ \bibnamefont {Raupach}}, \bibinfo
  {author} {\bibfnamefont {S.}~\bibnamefont {Gupta}}, \bibinfo {author}
  {\bibfnamefont {Z.}~\bibnamefont {Hadzibabic}},\ and\ \bibinfo {author}
  {\bibfnamefont {W.}~\bibnamefont {Ketterle}},\ }\bibfield  {title} {\bibinfo
  {title} {{Observation of Bose-Einstein Condensation of Molecules}},\ }\href
  {https://doi.org/10.1103/PhysRevLett.91.250401} {\bibfield  {journal}
  {\bibinfo  {journal} {Phys. Rev. Lett.}\ }\textbf {\bibinfo {volume} {91}},\
  \bibinfo {pages} {250401} (\bibinfo {year} {2003})}\BibitemShut {NoStop}%
\bibitem [{\citenamefont {Regal}\ \emph {et~al.}(2004)\citenamefont {Regal},
  \citenamefont {Greiner},\ and\ \citenamefont {Jin}}]{Regal2004oor}%
  \BibitemOpen
  \bibfield  {author} {\bibinfo {author} {\bibfnamefont {C.~A.}\ \bibnamefont
  {Regal}}, \bibinfo {author} {\bibfnamefont {M.}~\bibnamefont {Greiner}},\
  and\ \bibinfo {author} {\bibfnamefont {D.~S.}\ \bibnamefont {Jin}},\
  }\bibfield  {title} {\bibinfo {title} {{Observation of Resonance Condensation
  of Fermionic Atom Pairs}},\ }\href
  {https://doi.org/10.1103/PhysRevLett.92.040403} {\bibfield  {journal}
  {\bibinfo  {journal} {Phys. Rev. Lett.}\ }\textbf {\bibinfo {volume} {92}},\
  \bibinfo {pages} {040403} (\bibinfo {year} {2004})}\BibitemShut {NoStop}%
\bibitem [{\citenamefont {Zwierlein}\ \emph {et~al.}(2005)\citenamefont
  {Zwierlein}, \citenamefont {Abo-Shaeer}, \citenamefont {Schirotzek},
  \citenamefont {Schunck},\ and\ \citenamefont {Ketterle}}]{Zwierlein2005vas}%
  \BibitemOpen
  \bibfield  {author} {\bibinfo {author} {\bibfnamefont {M.~W.}\ \bibnamefont
  {Zwierlein}}, \bibinfo {author} {\bibfnamefont {J.~R.}\ \bibnamefont
  {Abo-Shaeer}}, \bibinfo {author} {\bibfnamefont {A.}~\bibnamefont
  {Schirotzek}}, \bibinfo {author} {\bibfnamefont {C.~H.}\ \bibnamefont
  {Schunck}},\ and\ \bibinfo {author} {\bibfnamefont {W.}~\bibnamefont
  {Ketterle}},\ }\bibfield  {title} {\bibinfo {title} {Vortices and
  superfluidity in a strongly interacting {F}ermi gas},\ }\href
  {https://doi.org/10.1038/nature03858} {\bibfield  {journal} {\bibinfo
  {journal} {Nature (London)}\ }\textbf {\bibinfo {volume} {435}},\ \bibinfo
  {pages} {1047} (\bibinfo {year} {2005})}\BibitemShut {NoStop}%
\bibitem [{\citenamefont {Zwerger}(2012)}]{Zwerger2012tbb}%
  \BibitemOpen
  \bibinfo {editor} {\bibfnamefont {W.}~\bibnamefont {Zwerger}},\ ed.,\
  \href@noop {} {\emph {\bibinfo {title} {The BCS-BEC Crossover and the Unitary
  Fermi Gas}}}\ (\bibinfo  {publisher} {Springer, Berlin Heidelberg},\ \bibinfo
  {year} {2012})\BibitemShut {NoStop}%
\bibitem [{\citenamefont {Strinati}\ \emph {et~al.}(2018)\citenamefont
  {Strinati}, \citenamefont {Pieri}, \citenamefont {R{\"o}pke}, \citenamefont
  {Schuck},\ and\ \citenamefont {Urban}}]{Strinati2018tbb}%
  \BibitemOpen
  \bibfield  {author} {\bibinfo {author} {\bibfnamefont {G.~C.}\ \bibnamefont
  {Strinati}}, \bibinfo {author} {\bibfnamefont {P.}~\bibnamefont {Pieri}},
  \bibinfo {author} {\bibfnamefont {G.}~\bibnamefont {R{\"o}pke}}, \bibinfo
  {author} {\bibfnamefont {P.}~\bibnamefont {Schuck}},\ and\ \bibinfo {author}
  {\bibfnamefont {M.}~\bibnamefont {Urban}},\ }\bibfield  {title} {\bibinfo
  {title} {The {BCS-BEC} crossover: From ultra-cold {F}ermi gases to nuclear
  systems},\ }\href {https://doi.org/10.1016/j.physrep.2018.02.004} {\bibfield
  {journal} {\bibinfo  {journal} {Phys. Rep.}\ }\textbf {\bibinfo {volume}
  {738}},\ \bibinfo {pages} {1} (\bibinfo {year} {2018})}\BibitemShut {NoStop}%
\bibitem [{\citenamefont {Ospelkaus}\ \emph {et~al.}(2006)\citenamefont
  {Ospelkaus}, \citenamefont {Ospelkaus}, \citenamefont {Humbert},
  \citenamefont {Ernst}, \citenamefont {Sengstock},\ and\ \citenamefont
  {Bongs}}]{Ospelkaus2006uhm}%
  \BibitemOpen
  \bibfield  {author} {\bibinfo {author} {\bibfnamefont {C.}~\bibnamefont
  {Ospelkaus}}, \bibinfo {author} {\bibfnamefont {S.}~\bibnamefont
  {Ospelkaus}}, \bibinfo {author} {\bibfnamefont {L.}~\bibnamefont {Humbert}},
  \bibinfo {author} {\bibfnamefont {P.}~\bibnamefont {Ernst}}, \bibinfo
  {author} {\bibfnamefont {K.}~\bibnamefont {Sengstock}},\ and\ \bibinfo
  {author} {\bibfnamefont {K.}~\bibnamefont {Bongs}},\ }\bibfield  {title}
  {\bibinfo {title} {{Ultracold Heteronuclear Molecules in a 3D Optical
  Lattice}},\ }\href {https://doi.org/10.1103/PhysRevLett.97.120402} {\bibfield
   {journal} {\bibinfo  {journal} {Phys. Rev. Lett.}\ }\textbf {\bibinfo
  {volume} {97}},\ \bibinfo {eid} {120402} (\bibinfo {year}
  {2006})}\BibitemShut {NoStop}%
\bibitem [{\citenamefont {Weber}\ \emph {et~al.}(2008)\citenamefont {Weber},
  \citenamefont {Barontini}, \citenamefont {Catani}, \citenamefont
  {Thalhammer}, \citenamefont {Inguscio},\ and\ \citenamefont
  {Minardi}}]{Weber2008aou}%
  \BibitemOpen
  \bibfield  {author} {\bibinfo {author} {\bibfnamefont {C.}~\bibnamefont
  {Weber}}, \bibinfo {author} {\bibfnamefont {G.}~\bibnamefont {Barontini}},
  \bibinfo {author} {\bibfnamefont {J.}~\bibnamefont {Catani}}, \bibinfo
  {author} {\bibfnamefont {G.}~\bibnamefont {Thalhammer}}, \bibinfo {author}
  {\bibfnamefont {M.}~\bibnamefont {Inguscio}},\ and\ \bibinfo {author}
  {\bibfnamefont {F.}~\bibnamefont {Minardi}},\ }\bibfield  {title} {\bibinfo
  {title} {Association of ultracold double-species bosonic molecules},\ }\href
  {https://doi.org/10.1103/PhysRevA.78.061601} {\bibfield  {journal} {\bibinfo
  {journal} {Phys. Rev. A}\ }\textbf {\bibinfo {volume} {78}},\ \bibinfo
  {pages} {061601(R)} (\bibinfo {year} {2008})}\BibitemShut {NoStop}%
\bibitem [{\citenamefont {Spiegelhalder}\ \emph {et~al.}(2010)\citenamefont
  {Spiegelhalder}, \citenamefont {Trenkwalder}, \citenamefont {Naik},
  \citenamefont {Kerner}, \citenamefont {Wille}, \citenamefont {Hendl},
  \citenamefont {Schreck},\ and\ \citenamefont {Grimm}}]{Spiegelhalder2010aop}%
  \BibitemOpen
  \bibfield  {author} {\bibinfo {author} {\bibfnamefont {F.~M.}\ \bibnamefont
  {Spiegelhalder}}, \bibinfo {author} {\bibfnamefont {A.}~\bibnamefont
  {Trenkwalder}}, \bibinfo {author} {\bibfnamefont {D.}~\bibnamefont {Naik}},
  \bibinfo {author} {\bibfnamefont {G.}~\bibnamefont {Kerner}}, \bibinfo
  {author} {\bibfnamefont {E.}~\bibnamefont {Wille}}, \bibinfo {author}
  {\bibfnamefont {G.}~\bibnamefont {Hendl}}, \bibinfo {author} {\bibfnamefont
  {F.}~\bibnamefont {Schreck}},\ and\ \bibinfo {author} {\bibfnamefont
  {R.}~\bibnamefont {Grimm}},\ }\bibfield  {title} {\bibinfo {title}
  {{All-optical production of a degenerate mixture of $^6$Li and $^{40}$K and
  creation of heteronuclear molecules}},\ }\href
  {https://doi.org/10.1103/PhysRevA.81.043637} {\bibfield  {journal} {\bibinfo
  {journal} {Phys. Rev. A}\ }\textbf {\bibinfo {volume} {81}},\ \bibinfo
  {pages} {043637} (\bibinfo {year} {2010})}\BibitemShut {NoStop}%
\bibitem [{\citenamefont {Wu}\ \emph {et~al.}(2012)\citenamefont {Wu},
  \citenamefont {Park}, \citenamefont {Ahmadi}, \citenamefont {Will},\ and\
  \citenamefont {Zwierlein}}]{Wu2012uff}%
  \BibitemOpen
  \bibfield  {author} {\bibinfo {author} {\bibfnamefont {C.-H.}\ \bibnamefont
  {Wu}}, \bibinfo {author} {\bibfnamefont {J.~W.}\ \bibnamefont {Park}},
  \bibinfo {author} {\bibfnamefont {P.}~\bibnamefont {Ahmadi}}, \bibinfo
  {author} {\bibfnamefont {S.}~\bibnamefont {Will}},\ and\ \bibinfo {author}
  {\bibfnamefont {M.~W.}\ \bibnamefont {Zwierlein}},\ }\bibfield  {title}
  {\bibinfo {title} {{Ultracold Fermionic Feshbach Molecules of
  $^{23}\mathrm{Na}^{40}\mathrm{K}$}},\ }\href
  {https://doi.org/10.1103/PhysRevLett.109.085301} {\bibfield  {journal}
  {\bibinfo  {journal} {Phys. Rev. Lett.}\ }\textbf {\bibinfo {volume} {109}},\
  \bibinfo {pages} {085301} (\bibinfo {year} {2012})}\BibitemShut {NoStop}%
\bibitem [{\citenamefont {Heo}\ \emph {et~al.}(2012)\citenamefont {Heo},
  \citenamefont {Wang}, \citenamefont {Christensen}, \citenamefont {Rvachov},
  \citenamefont {Cotta}, \citenamefont {Choi}, \citenamefont {Lee},\ and\
  \citenamefont {Ketterle}}]{Heo2012fou}%
  \BibitemOpen
  \bibfield  {author} {\bibinfo {author} {\bibfnamefont {M.-S.}\ \bibnamefont
  {Heo}}, \bibinfo {author} {\bibfnamefont {T.~T.}\ \bibnamefont {Wang}},
  \bibinfo {author} {\bibfnamefont {C.~A.}\ \bibnamefont {Christensen}},
  \bibinfo {author} {\bibfnamefont {T.~M.}\ \bibnamefont {Rvachov}}, \bibinfo
  {author} {\bibfnamefont {D.~A.}\ \bibnamefont {Cotta}}, \bibinfo {author}
  {\bibfnamefont {J.-H.}\ \bibnamefont {Choi}}, \bibinfo {author}
  {\bibfnamefont {Y.-R.}\ \bibnamefont {Lee}},\ and\ \bibinfo {author}
  {\bibfnamefont {W.}~\bibnamefont {Ketterle}},\ }\bibfield  {title} {\bibinfo
  {title} {{Formation of ultracold fermionic NaLi Feshbach molecules}},\ }\href
  {https://doi.org/10.1103/PhysRevA.86.021602} {\bibfield  {journal} {\bibinfo
  {journal} {Phys. Rev. A}\ }\textbf {\bibinfo {volume} {86}},\ \bibinfo
  {pages} {021602(R)} (\bibinfo {year} {2012})}\BibitemShut {NoStop}%
\bibitem [{\citenamefont {K\"oppinger}\ \emph {et~al.}(2014)\citenamefont
  {K\"oppinger}, \citenamefont {McCarron}, \citenamefont {Jenkin},
  \citenamefont {Molony}, \citenamefont {Cho}, \citenamefont {Cornish},
  \citenamefont {Le~Sueur}, \citenamefont {Blackley},\ and\ \citenamefont
  {Hutson}}]{Koeppinger2014poo}%
  \BibitemOpen
  \bibfield  {author} {\bibinfo {author} {\bibfnamefont {M.~P.}\ \bibnamefont
  {K\"oppinger}}, \bibinfo {author} {\bibfnamefont {D.~J.}\ \bibnamefont
  {McCarron}}, \bibinfo {author} {\bibfnamefont {D.~L.}\ \bibnamefont
  {Jenkin}}, \bibinfo {author} {\bibfnamefont {P.~K.}\ \bibnamefont {Molony}},
  \bibinfo {author} {\bibfnamefont {H.-W.}\ \bibnamefont {Cho}}, \bibinfo
  {author} {\bibfnamefont {S.~L.}\ \bibnamefont {Cornish}}, \bibinfo {author}
  {\bibfnamefont {C.~R.}\ \bibnamefont {Le~Sueur}}, \bibinfo {author}
  {\bibfnamefont {C.~L.}\ \bibnamefont {Blackley}},\ and\ \bibinfo {author}
  {\bibfnamefont {J.~M.}\ \bibnamefont {Hutson}},\ }\bibfield  {title}
  {\bibinfo {title} {Production of optically trapped $^{87}\mathrm{RbCs}$
  {F}eshbach molecules},\ }\href {https://doi.org/10.1103/PhysRevA.89.033604}
  {\bibfield  {journal} {\bibinfo  {journal} {Phys. Rev. A}\ }\textbf {\bibinfo
  {volume} {89}},\ \bibinfo {pages} {033604} (\bibinfo {year}
  {2014})}\BibitemShut {NoStop}%
\bibitem [{\citenamefont {Takekoshi}\ \emph {et~al.}(2014)\citenamefont
  {Takekoshi}, \citenamefont {Reichs\"ollner}, \citenamefont {Schindewolf},
  \citenamefont {Hutson}, \citenamefont {Le~Sueur}, \citenamefont {Dulieu},
  \citenamefont {Ferlaino}, \citenamefont {Grimm},\ and\ \citenamefont
  {N\"agerl}}]{Takekoshi2014uds}%
  \BibitemOpen
  \bibfield  {author} {\bibinfo {author} {\bibfnamefont {T.}~\bibnamefont
  {Takekoshi}}, \bibinfo {author} {\bibfnamefont {L.}~\bibnamefont
  {Reichs\"ollner}}, \bibinfo {author} {\bibfnamefont {A.}~\bibnamefont
  {Schindewolf}}, \bibinfo {author} {\bibfnamefont {J.~M.}\ \bibnamefont
  {Hutson}}, \bibinfo {author} {\bibfnamefont {C.~R.}\ \bibnamefont
  {Le~Sueur}}, \bibinfo {author} {\bibfnamefont {O.}~\bibnamefont {Dulieu}},
  \bibinfo {author} {\bibfnamefont {F.}~\bibnamefont {Ferlaino}}, \bibinfo
  {author} {\bibfnamefont {R.}~\bibnamefont {Grimm}},\ and\ \bibinfo {author}
  {\bibfnamefont {H.-C.}\ \bibnamefont {N\"agerl}},\ }\bibfield  {title}
  {\bibinfo {title} {{Ultracold Dense Samples of Dipolar RbCs Molecules in the
  Rovibrational and Hyperfine Ground State}},\ }\href
  {https://doi.org/10.1103/PhysRevLett.113.205301} {\bibfield  {journal}
  {\bibinfo  {journal} {Phys. Rev. Lett.}\ }\textbf {\bibinfo {volume} {113}},\
  \bibinfo {pages} {205301} (\bibinfo {year} {2014})}\BibitemShut {NoStop}%
\bibitem [{\citenamefont {Wang}\ \emph {et~al.}(2015)\citenamefont {Wang},
  \citenamefont {He}, \citenamefont {Li}, \citenamefont {Zhu}, \citenamefont
  {Chen},\ and\ \citenamefont {Wang}}]{Wang2015fou}%
  \BibitemOpen
  \bibfield  {author} {\bibinfo {author} {\bibfnamefont {F.}~\bibnamefont
  {Wang}}, \bibinfo {author} {\bibfnamefont {X.}~\bibnamefont {He}}, \bibinfo
  {author} {\bibfnamefont {X.}~\bibnamefont {Li}}, \bibinfo {author}
  {\bibfnamefont {B.}~\bibnamefont {Zhu}}, \bibinfo {author} {\bibfnamefont
  {J.}~\bibnamefont {Chen}},\ and\ \bibinfo {author} {\bibfnamefont
  {D.}~\bibnamefont {Wang}},\ }\bibfield  {title} {\bibinfo {title} {{Formation
  of ultracold NaRb Feshbach molecules}},\ }\href
  {https://doi.org/10.1088/1367-2630/17/3/035003} {\bibfield  {journal}
  {\bibinfo  {journal} {New J. Phys.}\ }\textbf {\bibinfo {volume} {17}},\
  \bibinfo {pages} {035003} (\bibinfo {year} {2015})}\BibitemShut {NoStop}%
\bibitem [{\citenamefont {Lam}\ \emph {et~al.}(2022)\citenamefont {Lam},
  \citenamefont {Bigagli}, \citenamefont {Warner}, \citenamefont {Yuan},
  \citenamefont {Zhang}, \citenamefont {Tiemann}, \citenamefont {Stevenson},\
  and\ \citenamefont {Will}}]{Lam2022HPS}%
  \BibitemOpen
  \bibfield  {author} {\bibinfo {author} {\bibfnamefont {A.~Z.}\ \bibnamefont
  {Lam}}, \bibinfo {author} {\bibfnamefont {N.}~\bibnamefont {Bigagli}},
  \bibinfo {author} {\bibfnamefont {C.}~\bibnamefont {Warner}}, \bibinfo
  {author} {\bibfnamefont {W.}~\bibnamefont {Yuan}}, \bibinfo {author}
  {\bibfnamefont {S.}~\bibnamefont {Zhang}}, \bibinfo {author} {\bibfnamefont
  {E.}~\bibnamefont {Tiemann}}, \bibinfo {author} {\bibfnamefont
  {I.}~\bibnamefont {Stevenson}},\ and\ \bibinfo {author} {\bibfnamefont
  {S.}~\bibnamefont {Will}},\ }\bibfield  {title} {\bibinfo {title} {{High
  phase-space density gas of NaCs Feshbach molecules}},\ }\href
  {https://doi.org/10.1103/PhysRevResearch.4.L022019} {\bibfield  {journal}
  {\bibinfo  {journal} {Phys. Rev. Research}\ }\textbf {\bibinfo {volume}
  {4}},\ \bibinfo {pages} {L022019} (\bibinfo {year} {2022})}\BibitemShut
  {NoStop}%
\bibitem [{\citenamefont {Taglieber}\ \emph {et~al.}(2008)\citenamefont
  {Taglieber}, \citenamefont {Voigt}, \citenamefont {Aoki}, \citenamefont
  {H\"{a}nsch},\ and\ \citenamefont {Dieckmann}}]{Taglieber2008qdt}%
  \BibitemOpen
  \bibfield  {author} {\bibinfo {author} {\bibfnamefont {M.}~\bibnamefont
  {Taglieber}}, \bibinfo {author} {\bibfnamefont {A.-C.}\ \bibnamefont
  {Voigt}}, \bibinfo {author} {\bibfnamefont {T.}~\bibnamefont {Aoki}},
  \bibinfo {author} {\bibfnamefont {T.~W.}\ \bibnamefont {H\"{a}nsch}},\ and\
  \bibinfo {author} {\bibfnamefont {K.}~\bibnamefont {Dieckmann}},\ }\bibfield
  {title} {\bibinfo {title} {{Quantum Degenerate Two-Species Fermi-Fermi
  Mixture Coexisting with a Bose-Einstein Condensate}},\ }\href
  {https://doi.org/10.1103/PhysRevLett.100.010401} {\bibfield  {journal}
  {\bibinfo  {journal} {Phys. Rev. Lett.}\ }\textbf {\bibinfo {volume} {100}},\
  \bibinfo {eid} {010401} (\bibinfo {year} {2008})}\BibitemShut {NoStop}%
\bibitem [{\citenamefont {Wille}\ \emph {et~al.}(2008)\citenamefont {Wille},
  \citenamefont {Spiegelhalder}, \citenamefont {Kerner}, \citenamefont {Naik},
  \citenamefont {Trenkwalder}, \citenamefont {Hendl}, \citenamefont {Schreck},
  \citenamefont {Grimm}, \citenamefont {Tiecke}, \citenamefont {Walraven},
  \citenamefont {Kokkelmans}, \citenamefont {Tiesinga},\ and\ \citenamefont
  {Julienne}}]{Wille2008eau}%
  \BibitemOpen
  \bibfield  {author} {\bibinfo {author} {\bibfnamefont {E.}~\bibnamefont
  {Wille}}, \bibinfo {author} {\bibfnamefont {F.~M.}\ \bibnamefont
  {Spiegelhalder}}, \bibinfo {author} {\bibfnamefont {G.}~\bibnamefont
  {Kerner}}, \bibinfo {author} {\bibfnamefont {D.}~\bibnamefont {Naik}},
  \bibinfo {author} {\bibfnamefont {A.}~\bibnamefont {Trenkwalder}}, \bibinfo
  {author} {\bibfnamefont {G.}~\bibnamefont {Hendl}}, \bibinfo {author}
  {\bibfnamefont {F.}~\bibnamefont {Schreck}}, \bibinfo {author} {\bibfnamefont
  {R.}~\bibnamefont {Grimm}}, \bibinfo {author} {\bibfnamefont {T.~G.}\
  \bibnamefont {Tiecke}}, \bibinfo {author} {\bibfnamefont {J.~T.~M.}\
  \bibnamefont {Walraven}}, \bibinfo {author} {\bibfnamefont {S.~J. J. M.~F.}\
  \bibnamefont {Kokkelmans}}, \bibinfo {author} {\bibfnamefont
  {E.}~\bibnamefont {Tiesinga}},\ and\ \bibinfo {author} {\bibfnamefont
  {P.~S.}\ \bibnamefont {Julienne}},\ }\bibfield  {title} {\bibinfo {title}
  {{Exploring an Ultracold Fermi-Fermi Mixture: Interspecies Feshbach
  Resonances and Scattering Properties of $^{6}\mathrm{Li}$ and
  $^{40}\mathrm{K}$}},\ }\href {https://doi.org/10.1103/PhysRevLett.100.053201}
  {\bibfield  {journal} {\bibinfo  {journal} {Phys. Rev. Lett.}\ }\textbf
  {\bibinfo {volume} {100}},\ \bibinfo {eid} {053201} (\bibinfo {year}
  {2008})}\BibitemShut {NoStop}%
\bibitem [{\citenamefont {Voigt}\ \emph {et~al.}(2009)\citenamefont {Voigt},
  \citenamefont {Taglieber}, \citenamefont {Costa}, \citenamefont {Aoki},
  \citenamefont {Wieser}, \citenamefont {H\"ansch},\ and\ \citenamefont
  {Dieckmann}}]{Voigt2009uhf}%
  \BibitemOpen
  \bibfield  {author} {\bibinfo {author} {\bibfnamefont {A.-C.}\ \bibnamefont
  {Voigt}}, \bibinfo {author} {\bibfnamefont {M.}~\bibnamefont {Taglieber}},
  \bibinfo {author} {\bibfnamefont {L.}~\bibnamefont {Costa}}, \bibinfo
  {author} {\bibfnamefont {T.}~\bibnamefont {Aoki}}, \bibinfo {author}
  {\bibfnamefont {W.}~\bibnamefont {Wieser}}, \bibinfo {author} {\bibfnamefont
  {T.~W.}\ \bibnamefont {H\"ansch}},\ and\ \bibinfo {author} {\bibfnamefont
  {K.}~\bibnamefont {Dieckmann}},\ }\bibfield  {title} {\bibinfo {title}
  {{Ultracold Heteronuclear Fermi-Fermi Molecules}},\ }\href
  {https://doi.org/10.1103/PhysRevLett.102.020405} {\bibfield  {journal}
  {\bibinfo  {journal} {Phys. Rev. Lett.}\ }\textbf {\bibinfo {volume} {102}},\
  \bibinfo {pages} {020405} (\bibinfo {year} {2009})},\ \bibinfo {note} {{\it
  ibid.} {\bf 105}, 269904(E) (2010).}\BibitemShut {Stop}%
\bibitem [{\citenamefont {Naik}\ \emph {et~al.}(2011)\citenamefont {Naik},
  \citenamefont {Trenkwalder}, \citenamefont {Kohstall}, \citenamefont
  {Spiegelhalder}, \citenamefont {Zaccanti}, \citenamefont {Hendl},
  \citenamefont {Schreck}, \citenamefont {Grimm}, \citenamefont {Hanna},\ and\
  \citenamefont {Julienne}}]{Naik2011fri}%
  \BibitemOpen
  \bibfield  {author} {\bibinfo {author} {\bibfnamefont {D.}~\bibnamefont
  {Naik}}, \bibinfo {author} {\bibfnamefont {A.}~\bibnamefont {Trenkwalder}},
  \bibinfo {author} {\bibfnamefont {C.}~\bibnamefont {Kohstall}}, \bibinfo
  {author} {\bibfnamefont {F.~M.}\ \bibnamefont {Spiegelhalder}}, \bibinfo
  {author} {\bibfnamefont {M.}~\bibnamefont {Zaccanti}}, \bibinfo {author}
  {\bibfnamefont {G.}~\bibnamefont {Hendl}}, \bibinfo {author} {\bibfnamefont
  {F.}~\bibnamefont {Schreck}}, \bibinfo {author} {\bibfnamefont
  {R.}~\bibnamefont {Grimm}}, \bibinfo {author} {\bibfnamefont
  {T.}~\bibnamefont {Hanna}},\ and\ \bibinfo {author} {\bibfnamefont
  {P.}~\bibnamefont {Julienne}},\ }\bibfield  {title} {\bibinfo {title}
  {{Feshbach resonances in the $^6$Li-$^{40}$K Fermi-Fermi mixture: Elastic
  versus inelastic interactions}},\ }\href
  {https://doi.org/10.1140/epjd/e2010-10591-2} {\bibfield  {journal} {\bibinfo
  {journal} {Eur. Phys. J. D}\ }\textbf {\bibinfo {volume} {65}},\ \bibinfo
  {pages} {55} (\bibinfo {year} {2011})}\BibitemShut {NoStop}%
\bibitem [{\citenamefont {Hara}\ \emph {et~al.}(2011)\citenamefont {Hara},
  \citenamefont {Takasu}, \citenamefont {Yamaoka}, \citenamefont {Doyle},\ and\
  \citenamefont {Takahashi}}]{Hara2011qdm}%
  \BibitemOpen
  \bibfield  {author} {\bibinfo {author} {\bibfnamefont {H.}~\bibnamefont
  {Hara}}, \bibinfo {author} {\bibfnamefont {Y.}~\bibnamefont {Takasu}},
  \bibinfo {author} {\bibfnamefont {Y.}~\bibnamefont {Yamaoka}}, \bibinfo
  {author} {\bibfnamefont {J.~M.}\ \bibnamefont {Doyle}},\ and\ \bibinfo
  {author} {\bibfnamefont {Y.}~\bibnamefont {Takahashi}},\ }\bibfield  {title}
  {\bibinfo {title} {{Q}uantum {D}egenerate {M}ixtures of {A}lkali and
  {A}lkaline-{E}arth-{L}ike {A}toms},\ }\href
  {https://doi.org/10.1103/PhysRevLett.106.205304} {\bibfield  {journal}
  {\bibinfo  {journal} {Phys. Rev. Lett.}\ }\textbf {\bibinfo {volume} {106}},\
  \bibinfo {pages} {205304} (\bibinfo {year} {2011})}\BibitemShut {NoStop}%
\bibitem [{\citenamefont {Green}\ \emph {et~al.}(2020)\citenamefont {Green},
  \citenamefont {Li}, \citenamefont {See~Toh}, \citenamefont {Tang},
  \citenamefont {McCormick}, \citenamefont {Li}, \citenamefont {Tiesinga},
  \citenamefont {Kotochigova},\ and\ \citenamefont {Gupta}}]{Green2020fri}%
  \BibitemOpen
  \bibfield  {author} {\bibinfo {author} {\bibfnamefont {A.}~\bibnamefont
  {Green}}, \bibinfo {author} {\bibfnamefont {H.}~\bibnamefont {Li}}, \bibinfo
  {author} {\bibfnamefont {J.~H.}\ \bibnamefont {See~Toh}}, \bibinfo {author}
  {\bibfnamefont {X.}~\bibnamefont {Tang}}, \bibinfo {author} {\bibfnamefont
  {K.~C.}\ \bibnamefont {McCormick}}, \bibinfo {author} {\bibfnamefont
  {M.}~\bibnamefont {Li}}, \bibinfo {author} {\bibfnamefont {E.}~\bibnamefont
  {Tiesinga}}, \bibinfo {author} {\bibfnamefont {S.}~\bibnamefont
  {Kotochigova}},\ and\ \bibinfo {author} {\bibfnamefont {S.}~\bibnamefont
  {Gupta}},\ }\bibfield  {title} {\bibinfo {title} {{F}eshbach {R}esonances in
  $p$-{W}ave {T}hree-{B}ody {R}ecombination within {F}ermi-{F}ermi {M}ixtures
  of {O}pen-{S}hell $^{6}\mathrm{Li}$ and {C}losed-{S}hell $^{173}\mathrm{Yb}$
  {A}toms},\ }\href {https://doi.org/10.1103/PhysRevX.10.031037} {\bibfield
  {journal} {\bibinfo  {journal} {Phys. Rev. X}\ }\textbf {\bibinfo {volume}
  {10}},\ \bibinfo {pages} {031037} (\bibinfo {year} {2020})}\BibitemShut
  {NoStop}%
\bibitem [{\citenamefont {Ravensbergen}\ \emph
  {et~al.}(2018{\natexlab{a}})\citenamefont {Ravensbergen}, \citenamefont
  {Corre}, \citenamefont {Soave}, \citenamefont {Kreyer}, \citenamefont
  {Kiri\-lov},\ and\ \citenamefont {Grimm}}]{Ravensbergen2018poa}%
  \BibitemOpen
  \bibfield  {author} {\bibinfo {author} {\bibfnamefont {C.}~\bibnamefont
  {Ravensbergen}}, \bibinfo {author} {\bibfnamefont {V.}~\bibnamefont {Corre}},
  \bibinfo {author} {\bibfnamefont {E.}~\bibnamefont {Soave}}, \bibinfo
  {author} {\bibfnamefont {M.}~\bibnamefont {Kreyer}}, \bibinfo {author}
  {\bibfnamefont {E.}~\bibnamefont {Kiri\-lov}},\ and\ \bibinfo {author}
  {\bibfnamefont {R.}~\bibnamefont {Grimm}},\ }\bibfield  {title} {\bibinfo
  {title} {Production of a degenerate {F}ermi-{F}ermi mixture of dysprosium and
  potassium atoms},\ }\href {https://doi.org/10.1103/PhysRevA.98.063624}
  {\bibfield  {journal} {\bibinfo  {journal} {Phys. Rev. A}\ }\textbf {\bibinfo
  {volume} {98}},\ \bibinfo {pages} {063624} (\bibinfo {year}
  {2018}{\natexlab{a}})},\ \bibinfo {note} {{\it ibid.} {\bf 101}, 059903(E)
  (2020).}\BibitemShut {Stop}%
\bibitem [{\citenamefont {Ravensbergen}\ \emph {et~al.}(2020)\citenamefont
  {Ravensbergen}, \citenamefont {Soave}, \citenamefont {Corre}, \citenamefont
  {Kreyer}, \citenamefont {Huang}, \citenamefont {Kirilov},\ and\ \citenamefont
  {Grimm}}]{Ravensbergen2020rif}%
  \BibitemOpen
  \bibfield  {author} {\bibinfo {author} {\bibfnamefont {C.}~\bibnamefont
  {Ravensbergen}}, \bibinfo {author} {\bibfnamefont {E.}~\bibnamefont {Soave}},
  \bibinfo {author} {\bibfnamefont {V.}~\bibnamefont {Corre}}, \bibinfo
  {author} {\bibfnamefont {M.}~\bibnamefont {Kreyer}}, \bibinfo {author}
  {\bibfnamefont {B.}~\bibnamefont {Huang}}, \bibinfo {author} {\bibfnamefont
  {E.}~\bibnamefont {Kirilov}},\ and\ \bibinfo {author} {\bibfnamefont
  {R.}~\bibnamefont {Grimm}},\ }\bibfield  {title} {\bibinfo {title}
  {{Resonantly Interacting Fermi-Fermi Mixture of $^{161}\mathrm{Dy}$ and
  $^{40}\mathrm{K}$}},\ }\href {https://doi.org/10.1103/PhysRevLett.124.203402}
  {\bibfield  {journal} {\bibinfo  {journal} {Phys. Rev. Lett.}\ }\textbf
  {\bibinfo {volume} {124}},\ \bibinfo {pages} {203402} (\bibinfo {year}
  {2020})}\BibitemShut {NoStop}%
\bibitem [{\citenamefont {Ciamei}\ \emph
  {et~al.}(2022{\natexlab{a}})\citenamefont {Ciamei}, \citenamefont {Finelli},
  \citenamefont {Cosco}, \citenamefont {Inguscio}, \citenamefont
  {Trenkwalder},\ and\ \citenamefont {Zaccanti}}]{Ciamei2022ddf}%
  \BibitemOpen
  \bibfield  {author} {\bibinfo {author} {\bibfnamefont {A.}~\bibnamefont
  {Ciamei}}, \bibinfo {author} {\bibfnamefont {S.}~\bibnamefont {Finelli}},
  \bibinfo {author} {\bibfnamefont {A.}~\bibnamefont {Cosco}}, \bibinfo
  {author} {\bibfnamefont {M.}~\bibnamefont {Inguscio}}, \bibinfo {author}
  {\bibfnamefont {A.}~\bibnamefont {Trenkwalder}},\ and\ \bibinfo {author}
  {\bibfnamefont {M.}~\bibnamefont {Zaccanti}},\ }\bibfield  {title} {\bibinfo
  {title} {Double-degenerate fermi mixtures of $^{6}\mathrm{Li}$ and
  $^{53}\mathrm{Cr}$ atoms},\ }\href
  {https://doi.org/10.1103/PhysRevA.106.053318} {\bibfield  {journal} {\bibinfo
   {journal} {Phys. Rev. A}\ }\textbf {\bibinfo {volume} {106}},\ \bibinfo
  {pages} {053318} (\bibinfo {year} {2022}{\natexlab{a}})}\BibitemShut
  {NoStop}%
\bibitem [{\citenamefont {Ciamei}\ \emph
  {et~al.}(2022{\natexlab{b}})\citenamefont {Ciamei}, \citenamefont {Finelli},
  \citenamefont {Trenkwalder}, \citenamefont {Inguscio}, \citenamefont
  {Simoni},\ and\ \citenamefont {Zaccanti}}]{Ciamei2022euc}%
  \BibitemOpen
  \bibfield  {author} {\bibinfo {author} {\bibfnamefont {A.}~\bibnamefont
  {Ciamei}}, \bibinfo {author} {\bibfnamefont {S.}~\bibnamefont {Finelli}},
  \bibinfo {author} {\bibfnamefont {A.}~\bibnamefont {Trenkwalder}}, \bibinfo
  {author} {\bibfnamefont {M.}~\bibnamefont {Inguscio}}, \bibinfo {author}
  {\bibfnamefont {A.}~\bibnamefont {Simoni}},\ and\ \bibinfo {author}
  {\bibfnamefont {M.}~\bibnamefont {Zaccanti}},\ }\bibfield  {title} {\bibinfo
  {title} {{Exploring Ultracold Collisions in
  $^{6}\mathrm{Li}\text{\ensuremath{-}}^{53}\mathrm{Cr}$ Fermi Mixtures:
  Feshbach Resonances and Scattering Properties of a Novel Alkali-Transition
  Metal System}},\ }\href {https://doi.org/10.1103/PhysRevLett.129.093402}
  {\bibfield  {journal} {\bibinfo  {journal} {Phys. Rev. Lett.}\ }\textbf
  {\bibinfo {volume} {129}},\ \bibinfo {pages} {093402} (\bibinfo {year}
  {2022}{\natexlab{b}})}\BibitemShut {NoStop}%
\bibitem [{\citenamefont {Gubbels}\ and\ \citenamefont
  {Stoof}(2013)}]{Gubbels2013ifg}%
  \BibitemOpen
  \bibfield  {author} {\bibinfo {author} {\bibfnamefont {K.~B.}\ \bibnamefont
  {Gubbels}}\ and\ \bibinfo {author} {\bibfnamefont {H.~T.~C.}\ \bibnamefont
  {Stoof}},\ }\bibfield  {title} {\bibinfo {title} {Imbalanced {F}ermi gases at
  unitarity},\ }\href {https://doi.org/10.1016/j.physrep.2012.11.004}
  {\bibfield  {journal} {\bibinfo  {journal} {Phys. Rep.}\ }\textbf {\bibinfo
  {volume} {525}},\ \bibinfo {pages} {255 } (\bibinfo {year}
  {2013})}\BibitemShut {NoStop}%
\bibitem [{\citenamefont {Gubbels}\ \emph {et~al.}(2009)\citenamefont
  {Gubbels}, \citenamefont {Baarsma},\ and\ \citenamefont
  {Stoof}}]{Gubbels2009lpi}%
  \BibitemOpen
  \bibfield  {author} {\bibinfo {author} {\bibfnamefont {K.~B.}\ \bibnamefont
  {Gubbels}}, \bibinfo {author} {\bibfnamefont {J.~E.}\ \bibnamefont
  {Baarsma}},\ and\ \bibinfo {author} {\bibfnamefont {H.~T.~C.}\ \bibnamefont
  {Stoof}},\ }\bibfield  {title} {\bibinfo {title} {Lifshitz {P}oint in the
  {P}hase {D}iagram of {R}esonantly {I}nteracting
  $^{6}\mathrm{Li}\mathrm{\text{-}}^{40}\mathrm{K}$ {M}ixtures},\ }\href
  {https://doi.org/10.1103/PhysRevLett.103.195301} {\bibfield  {journal}
  {\bibinfo  {journal} {Phys. Rev. Lett.}\ }\textbf {\bibinfo {volume} {103}},\
  \bibinfo {pages} {195301} (\bibinfo {year} {2009})}\BibitemShut {NoStop}%
\bibitem [{\citenamefont {Wang}\ \emph {et~al.}(2017)\citenamefont {Wang},
  \citenamefont {Che}, \citenamefont {Zhang},\ and\ \citenamefont
  {Chen}}]{Wang2017eeo}%
  \BibitemOpen
  \bibfield  {author} {\bibinfo {author} {\bibfnamefont {J.}~\bibnamefont
  {Wang}}, \bibinfo {author} {\bibfnamefont {Y.}~\bibnamefont {Che}}, \bibinfo
  {author} {\bibfnamefont {L.}~\bibnamefont {Zhang}},\ and\ \bibinfo {author}
  {\bibfnamefont {Q.}~\bibnamefont {Chen}},\ }\bibfield  {title} {\bibinfo
  {title} {Enhancement effect of mass imbalance on
  {Fulde-Ferrell-Larkin-Ovchinnikov} type of pairing in {Fermi-Fermi} mixtures
  of ultracold quantum gases},\ }\href {https://doi.org/10.1038/srep39783}
  {\bibfield  {journal} {\bibinfo  {journal} {Sci. Rep.}\ }\textbf {\bibinfo
  {volume} {7}},\ \bibinfo {pages} {39783} (\bibinfo {year}
  {2017})}\BibitemShut {NoStop}%
\bibitem [{\citenamefont {Pini}\ \emph {et~al.}(2021)\citenamefont {Pini},
  \citenamefont {Pieri}, \citenamefont {Grimm},\ and\ \citenamefont
  {Strinati}}]{Pini2021bmf}%
  \BibitemOpen
  \bibfield  {author} {\bibinfo {author} {\bibfnamefont {M.}~\bibnamefont
  {Pini}}, \bibinfo {author} {\bibfnamefont {P.}~\bibnamefont {Pieri}},
  \bibinfo {author} {\bibfnamefont {R.}~\bibnamefont {Grimm}},\ and\ \bibinfo
  {author} {\bibfnamefont {G.~C.}\ \bibnamefont {Strinati}},\ }\bibfield
  {title} {\bibinfo {title} {{Beyond-mean-field description of a trapped
  unitary Fermi gas with mass and population imbalance}},\ }\href
  {https://doi.org/10.1103/PhysRevA.103.023314} {\bibfield  {journal} {\bibinfo
   {journal} {Phys. Rev. A}\ }\textbf {\bibinfo {volume} {103}},\ \bibinfo
  {pages} {023314} (\bibinfo {year} {2021})}\BibitemShut {NoStop}%
\bibitem [{\citenamefont {Fulde}\ and\ \citenamefont
  {Ferrell}(1964)}]{Fulde1964sia}%
  \BibitemOpen
  \bibfield  {author} {\bibinfo {author} {\bibfnamefont {P.}~\bibnamefont
  {Fulde}}\ and\ \bibinfo {author} {\bibfnamefont {R.~A.}\ \bibnamefont
  {Ferrell}},\ }\bibfield  {title} {\bibinfo {title} {{Superconductivity in a
  Strong Spin-Exchange Field}},\ }\href
  {https://doi.org/10.1103/PhysRev.135.A550} {\bibfield  {journal} {\bibinfo
  {journal} {Phys. Rev.}\ }\textbf {\bibinfo {volume} {135}},\ \bibinfo {pages}
  {A550} (\bibinfo {year} {1964})}\BibitemShut {NoStop}%
\bibitem [{\citenamefont {Larkin}\ and\ \citenamefont
  {Ovchinnikov}(1964)}]{Larkin1964nss}%
  \BibitemOpen
  \bibfield  {author} {\bibinfo {author} {\bibfnamefont {A.~I.}\ \bibnamefont
  {Larkin}}\ and\ \bibinfo {author} {\bibfnamefont {Y.~N.}\ \bibnamefont
  {Ovchinnikov}},\ }\bibfield  {title} {\bibinfo {title} {Neodnorodnoye
  sostoyanie sverkhprovodnikov},\ }\href@noop {} {\bibfield  {journal}
  {\bibinfo  {journal} {Zh. Eksp. Teor. Fiz.}\ }\textbf {\bibinfo {volume}
  {47}},\ \bibinfo {pages} {1136} (\bibinfo {year} {1964})},\ \bibinfo {note}
  {[Sov. Phys. JETP 20, 762 (1965)].}\BibitemShut {Stop}%
\bibitem [{\citenamefont {Radzihovsky}\ and\ \citenamefont
  {Sheehy}(2010)}]{Radzihovsky2010ifr}%
  \BibitemOpen
  \bibfield  {author} {\bibinfo {author} {\bibfnamefont {L.}~\bibnamefont
  {Radzihovsky}}\ and\ \bibinfo {author} {\bibfnamefont {D.~E.}\ \bibnamefont
  {Sheehy}},\ }\bibfield  {title} {\bibinfo {title} {{Imbalanced
  Feshbach-resonant Fermi gases}},\ }\href
  {https://doi.org/10.1088/0034-4885/73/7/076501} {\bibfield  {journal}
  {\bibinfo  {journal} {Rep. Prog. Phys.}\ }\textbf {\bibinfo {volume} {73}},\
  \bibinfo {pages} {076501} (\bibinfo {year} {2010})}\BibitemShut {NoStop}%
\bibitem [{\citenamefont {Naidon}\ and\ \citenamefont
  {Endo}(2017)}]{Naidon2017epa}%
  \BibitemOpen
  \bibfield  {author} {\bibinfo {author} {\bibfnamefont {P.}~\bibnamefont
  {Naidon}}\ and\ \bibinfo {author} {\bibfnamefont {S.}~\bibnamefont {Endo}},\
  }\bibfield  {title} {\bibinfo {title} {Efimov physics: a review},\ }\href
  {https://doi.org/10.1088/1361-6633/aa50e8} {\bibfield  {journal} {\bibinfo
  {journal} {Rep. Prog. Phys.}\ }\textbf {\bibinfo {volume} {80}},\ \bibinfo
  {pages} {056001} (\bibinfo {year} {2017})}\BibitemShut {NoStop}%
\bibitem [{\citenamefont {Kartavtsev}\ and\ \citenamefont
  {Malykh}(2007)}]{Kartavtsev2007let}%
  \BibitemOpen
  \bibfield  {author} {\bibinfo {author} {\bibfnamefont {O.~I.}\ \bibnamefont
  {Kartavtsev}}\ and\ \bibinfo {author} {\bibfnamefont {A.~V.}\ \bibnamefont
  {Malykh}},\ }\bibfield  {title} {\bibinfo {title} {Low-energy three-body
  dynamics in binary quantum gases},\ }\href@noop {} {\bibfield  {journal}
  {\bibinfo  {journal} {J. Phys. B}\ }\textbf {\bibinfo {volume} {40}},\
  \bibinfo {pages} {1429} (\bibinfo {year} {2007})}\BibitemShut {NoStop}%
\bibitem [{\citenamefont {Ye}\ \emph {et~al.}(2022)\citenamefont {Ye},
  \citenamefont {Canali}, \citenamefont {Soave}, \citenamefont {Kreyer},
  \citenamefont {Yudkin}, \citenamefont {Ravensbergen}, \citenamefont
  {Kirilov},\ and\ \citenamefont {Grimm}}]{Ye2022OOL}%
  \BibitemOpen
  \bibfield  {author} {\bibinfo {author} {\bibfnamefont {Z.-X.}\ \bibnamefont
  {Ye}}, \bibinfo {author} {\bibfnamefont {A.}~\bibnamefont {Canali}}, \bibinfo
  {author} {\bibfnamefont {E.}~\bibnamefont {Soave}}, \bibinfo {author}
  {\bibfnamefont {M.}~\bibnamefont {Kreyer}}, \bibinfo {author} {\bibfnamefont
  {Y.}~\bibnamefont {Yudkin}}, \bibinfo {author} {\bibfnamefont
  {C.}~\bibnamefont {Ravensbergen}}, \bibinfo {author} {\bibfnamefont
  {E.}~\bibnamefont {Kirilov}},\ and\ \bibinfo {author} {\bibfnamefont
  {R.}~\bibnamefont {Grimm}},\ }\bibfield  {title} {\bibinfo {title}
  {{Observation of low-field Feshbach resonances between $^{161}\mathrm{Dy}$
  and $^{40}\mathrm{K}$}},\ }\href
  {https://doi.org/10.1103/PhysRevA.106.043314} {\bibfield  {journal} {\bibinfo
   {journal} {Phys. Rev. A}\ }\textbf {\bibinfo {volume} {106}},\ \bibinfo
  {pages} {043314} (\bibinfo {year} {2022})}\BibitemShut {NoStop}%
\bibitem [{\citenamefont {Lous}\ \emph {et~al.}(2017)\citenamefont {Lous},
  \citenamefont {Fritsche}, \citenamefont {Jag}, \citenamefont {Huang},\ and\
  \citenamefont {Grimm}}]{Lous2017toa}%
  \BibitemOpen
  \bibfield  {author} {\bibinfo {author} {\bibfnamefont {R.~S.}\ \bibnamefont
  {Lous}}, \bibinfo {author} {\bibfnamefont {I.}~\bibnamefont {Fritsche}},
  \bibinfo {author} {\bibfnamefont {M.}~\bibnamefont {Jag}}, \bibinfo {author}
  {\bibfnamefont {B.}~\bibnamefont {Huang}},\ and\ \bibinfo {author}
  {\bibfnamefont {R.}~\bibnamefont {Grimm}},\ }\bibfield  {title} {\bibinfo
  {title} {Thermometry of a deeply degenerate {F}ermi gas with a
  {B}ose-{E}instein condensate},\ }\href
  {https://doi.org/10.1103/PhysRevA.95.053627} {\bibfield  {journal} {\bibinfo
  {journal} {Phys. Rev. A}\ }\textbf {\bibinfo {volume} {95}},\ \bibinfo
  {pages} {053627} (\bibinfo {year} {2017})}\BibitemShut {NoStop}%
\bibitem [{\citenamefont {Ravensbergen}\ \emph
  {et~al.}(2018{\natexlab{b}})\citenamefont {Ravensbergen}, \citenamefont
  {Corre}, \citenamefont {Soave}, \citenamefont {Kreyer}, \citenamefont
  {Tzanova}, \citenamefont {Kiri\-lov},\ and\ \citenamefont
  {Grimm}}]{Ravensbergen2018ado}%
  \BibitemOpen
  \bibfield  {author} {\bibinfo {author} {\bibfnamefont {C.}~\bibnamefont
  {Ravensbergen}}, \bibinfo {author} {\bibfnamefont {V.}~\bibnamefont {Corre}},
  \bibinfo {author} {\bibfnamefont {E.}~\bibnamefont {Soave}}, \bibinfo
  {author} {\bibfnamefont {M.}~\bibnamefont {Kreyer}}, \bibinfo {author}
  {\bibfnamefont {S.}~\bibnamefont {Tzanova}}, \bibinfo {author} {\bibfnamefont
  {E.}~\bibnamefont {Kiri\-lov}},\ and\ \bibinfo {author} {\bibfnamefont
  {R.}~\bibnamefont {Grimm}},\ }\bibfield  {title} {\bibinfo {title} {{Accurate
  Determination of the Dynamical Polarizability of Dysprosium}},\ }\href
  {https://doi.org/10.1103/PhysRevLett.120.223001} {\bibfield  {journal}
  {\bibinfo  {journal} {Phys. Rev. Lett.}\ }\textbf {\bibinfo {volume} {120}},\
  \bibinfo {pages} {223001} (\bibinfo {year} {2018}{\natexlab{b}})}\BibitemShut
  {NoStop}%
\bibitem [{Note1()}]{Note1}%
  \BibitemOpen
  \bibinfo {note} {See Supplemental Material for the data files of all
  figures}\BibitemShut {NoStop}%
\bibitem [{\citenamefont {D{\"u}rr}\ \emph
  {et~al.}(2004{\natexlab{b}})\citenamefont {D{\"u}rr}, \citenamefont {Volz},\
  and\ \citenamefont {Rempe}}]{Duerr2004dou}%
  \BibitemOpen
  \bibfield  {author} {\bibinfo {author} {\bibfnamefont {S.}~\bibnamefont
  {D{\"u}rr}}, \bibinfo {author} {\bibfnamefont {T.}~\bibnamefont {Volz}},\
  and\ \bibinfo {author} {\bibfnamefont {G.}~\bibnamefont {Rempe}},\ }\bibfield
   {title} {\bibinfo {title} {Dissociation of ultracold molecules with
  {F}eshbach resonances},\ }\href {https://doi.org/10.1103/PhysRevA.70.031601}
  {\bibfield  {journal} {\bibinfo  {journal} {Phys. Rev. A}\ }\textbf {\bibinfo
  {volume} {70}},\ \bibinfo {pages} {031601(R)} (\bibinfo {year}
  {2004}{\natexlab{b}})}\BibitemShut {NoStop}%
\bibitem [{\citenamefont {De~Marco}\ \emph {et~al.}(2019)\citenamefont
  {De~Marco}, \citenamefont {Valtolina}, \citenamefont {Matsuda}, \citenamefont
  {Tobias}, \citenamefont {Covey},\ and\ \citenamefont {Ye}}]{Demarco2019adf}%
  \BibitemOpen
  \bibfield  {author} {\bibinfo {author} {\bibfnamefont {L.}~\bibnamefont
  {De~Marco}}, \bibinfo {author} {\bibfnamefont {G.}~\bibnamefont {Valtolina}},
  \bibinfo {author} {\bibfnamefont {K.}~\bibnamefont {Matsuda}}, \bibinfo
  {author} {\bibfnamefont {W.~G.}\ \bibnamefont {Tobias}}, \bibinfo {author}
  {\bibfnamefont {J.~P.}\ \bibnamefont {Covey}},\ and\ \bibinfo {author}
  {\bibfnamefont {J.}~\bibnamefont {Ye}},\ }\bibfield  {title} {\bibinfo
  {title} {{A degenerate Fermi gas of polar molecules}},\ }\href
  {https://doi.org/10.1126/science.aau7230} {\bibfield  {journal} {\bibinfo
  {journal} {Science}\ }\textbf {\bibinfo {volume} {363}},\ \bibinfo {pages}
  {853} (\bibinfo {year} {2019})}\BibitemShut {NoStop}%
\bibitem [{\citenamefont {Duda}\ \emph {et~al.}(2023)\citenamefont {Duda},
  \citenamefont {Chen}, \citenamefont {Schindewolf}, \citenamefont {Bause},
  \citenamefont {von Milczewski}, \citenamefont {Schmidt}, \citenamefont
  {Bloch},\ and\ \citenamefont {Luo}}]{Duda2023tfa}%
  \BibitemOpen
  \bibfield  {author} {\bibinfo {author} {\bibfnamefont {M.}~\bibnamefont
  {Duda}}, \bibinfo {author} {\bibfnamefont {X.-Y.}\ \bibnamefont {Chen}},
  \bibinfo {author} {\bibfnamefont {A.}~\bibnamefont {Schindewolf}}, \bibinfo
  {author} {\bibfnamefont {R.}~\bibnamefont {Bause}}, \bibinfo {author}
  {\bibfnamefont {J.}~\bibnamefont {von Milczewski}}, \bibinfo {author}
  {\bibfnamefont {R.}~\bibnamefont {Schmidt}}, \bibinfo {author} {\bibfnamefont
  {I.}~\bibnamefont {Bloch}},\ and\ \bibinfo {author} {\bibfnamefont {X.-Y.}\
  \bibnamefont {Luo}},\ }\bibfield  {title} {\bibinfo {title} {Transition from
  a polaronic condensate to a degenerate {F}ermi gas of heteronuclear
  molecules},\ }\href@noop {} {\bibfield  {journal} {\bibinfo  {journal} {Nat.
  Phys.}\ } (\bibinfo {year} {2023})}\BibitemShut {NoStop}%
\bibitem [{\citenamefont {Petrov}(2004)}]{Petrov2004TBP}%
  \BibitemOpen
  \bibfield  {author} {\bibinfo {author} {\bibfnamefont {D.~S.}\ \bibnamefont
  {Petrov}},\ }\bibfield  {title} {\bibinfo {title} {{Three-Boson Problem near
  a Narrow {F}eshbach Resonance}},\ }\href
  {https://doi.org/10.1103/PhysRevLett.93.143201} {\bibfield  {journal}
  {\bibinfo  {journal} {Phys. Rev. Lett.}\ }\textbf {\bibinfo {volume} {93}},\
  \bibinfo {pages} {143201} (\bibinfo {year} {2004})}\BibitemShut {NoStop}%
\bibitem [{\citenamefont {Braaten}\ and\ \citenamefont
  {Hammer}(2006)}]{Braaten2006UIF}%
  \BibitemOpen
  \bibfield  {author} {\bibinfo {author} {\bibfnamefont {E.}~\bibnamefont
  {Braaten}}\ and\ \bibinfo {author} {\bibfnamefont {H.-W.}\ \bibnamefont
  {Hammer}},\ }\bibfield  {title} {\bibinfo {title} {Universality in few-body
  systems with large scattering length},\ }\href@noop {} {\bibfield  {journal}
  {\bibinfo  {journal} {Phys. Rep.}\ }\textbf {\bibinfo {volume} {428}},\
  \bibinfo {pages} {259} (\bibinfo {year} {2006})}\BibitemShut {NoStop}%
\bibitem [{\citenamefont {Bloch}\ \emph {et~al.}(2008)\citenamefont {Bloch},
  \citenamefont {Dalibard},\ and\ \citenamefont {Zwerger}}]{Bloch2008MBP}%
  \BibitemOpen
  \bibfield  {author} {\bibinfo {author} {\bibfnamefont {I.}~\bibnamefont
  {Bloch}}, \bibinfo {author} {\bibfnamefont {J.}~\bibnamefont {Dalibard}},\
  and\ \bibinfo {author} {\bibfnamefont {W.}~\bibnamefont {Zwerger}},\
  }\bibfield  {title} {\bibinfo {title} {Many-body physics with ultracold
  gases},\ }\href {https://doi.org/10.1103/RevModPhys.80.885} {\bibfield
  {journal} {\bibinfo  {journal} {Rev. Mod. Phys.}\ }\textbf {\bibinfo {volume}
  {80}},\ \bibinfo {pages} {885} (\bibinfo {year} {2008})}\BibitemShut
  {NoStop}%
\bibitem [{\citenamefont {Lous}\ \emph {et~al.}(2018)\citenamefont {Lous},
  \citenamefont {Fritsche}, \citenamefont {Jag}, \citenamefont {Lehmann},
  \citenamefont {Kirilov}, \citenamefont {Huang},\ and\ \citenamefont
  {Grimm}}]{Lous2018PTI}%
  \BibitemOpen
  \bibfield  {author} {\bibinfo {author} {\bibfnamefont {R.~S.}\ \bibnamefont
  {Lous}}, \bibinfo {author} {\bibfnamefont {I.}~\bibnamefont {Fritsche}},
  \bibinfo {author} {\bibfnamefont {M.}~\bibnamefont {Jag}}, \bibinfo {author}
  {\bibfnamefont {F.}~\bibnamefont {Lehmann}}, \bibinfo {author} {\bibfnamefont
  {E.}~\bibnamefont {Kirilov}}, \bibinfo {author} {\bibfnamefont
  {B.}~\bibnamefont {Huang}},\ and\ \bibinfo {author} {\bibfnamefont
  {R.}~\bibnamefont {Grimm}},\ }\bibfield  {title} {\bibinfo {title} {{Probing
  the Interface of a Phase-Separated State in a Repulsive Bose-Fermi
  Mixture}},\ }\href {https://doi.org/10.1103/PhysRevLett.120.243403}
  {\bibfield  {journal} {\bibinfo  {journal} {Phys. Rev. Lett.}\ }\textbf
  {\bibinfo {volume} {120}},\ \bibinfo {pages} {243403} (\bibinfo {year}
  {2018})}\BibitemShut {NoStop}%
\bibitem [{\citenamefont {Thompson}\ \emph {et~al.}(2005)\citenamefont
  {Thompson}, \citenamefont {Hodby},\ and\ \citenamefont
  {Wieman}}]{Thompson2005UMP}%
  \BibitemOpen
  \bibfield  {author} {\bibinfo {author} {\bibfnamefont {S.~T.}\ \bibnamefont
  {Thompson}}, \bibinfo {author} {\bibfnamefont {E.}~\bibnamefont {Hodby}},\
  and\ \bibinfo {author} {\bibfnamefont {C.~E.}\ \bibnamefont {Wieman}},\
  }\bibfield  {title} {\bibinfo {title} {Ultracold molecule production via a
  resonant oscillating magnetic field},\ }\href@noop {} {\bibfield  {journal}
  {\bibinfo  {journal} {Phys. Rev. Lett.}\ }\textbf {\bibinfo {volume} {95}},\
  \bibinfo {eid} {190404} (\bibinfo {year} {2005})}\BibitemShut {NoStop}%
\bibitem [{\citenamefont {Claussen}\ \emph {et~al.}(2003)\citenamefont
  {Claussen}, \citenamefont {Kokkelmans}, \citenamefont {Thompson},
  \citenamefont {Donley}, \citenamefont {Hodby},\ and\ \citenamefont
  {Wieman}}]{Claussen2003VHP}%
  \BibitemOpen
  \bibfield  {author} {\bibinfo {author} {\bibfnamefont {N.~R.}\ \bibnamefont
  {Claussen}}, \bibinfo {author} {\bibfnamefont {S.~J. J. M.~F.}\ \bibnamefont
  {Kokkelmans}}, \bibinfo {author} {\bibfnamefont {S.~T.}\ \bibnamefont
  {Thompson}}, \bibinfo {author} {\bibfnamefont {E.~A.}\ \bibnamefont
  {Donley}}, \bibinfo {author} {\bibfnamefont {E.}~\bibnamefont {Hodby}},\ and\
  \bibinfo {author} {\bibfnamefont {C.~E.}\ \bibnamefont {Wieman}},\ }\bibfield
   {title} {\bibinfo {title} {Very-high-precision bound-state spectroscopy near
  a $^{85}${R}b {F}eshbach resonance},\ }\href
  {https://doi.org/10.1103/PhysRevA.67.060701} {\bibfield  {journal} {\bibinfo
  {journal} {Phys. Rev. A}\ }\textbf {\bibinfo {volume} {67}},\ \bibinfo {eid}
  {060701} (\bibinfo {year} {2003})}\BibitemShut {NoStop}%
\bibitem [{\citenamefont {Hanna}\ \emph {et~al.}(2007)\citenamefont {Hanna},
  \citenamefont {K\"{o}hler},\ and\ \citenamefont {Burnett}}]{Hanna2007AOM}%
  \BibitemOpen
  \bibfield  {author} {\bibinfo {author} {\bibfnamefont {T.~M.}\ \bibnamefont
  {Hanna}}, \bibinfo {author} {\bibfnamefont {T.}~\bibnamefont {K\"{o}hler}},\
  and\ \bibinfo {author} {\bibfnamefont {K.}~\bibnamefont {Burnett}},\
  }\bibfield  {title} {\bibinfo {title} {Association of molecules using a
  resonantly modulated magnetic field},\ }\href@noop {} {\bibfield  {journal}
  {\bibinfo  {journal} {Phys. Rev. A}\ }\textbf {\bibinfo {volume} {75}},\
  \bibinfo {eid} {013606} (\bibinfo {year} {2007})}\BibitemShut {NoStop}%
\bibitem [{\citenamefont {Mohapatra}\ and\ \citenamefont
  {Braaten}(2015)}]{Mohapatra2015HAS}%
  \BibitemOpen
  \bibfield  {author} {\bibinfo {author} {\bibfnamefont {A.}~\bibnamefont
  {Mohapatra}}\ and\ \bibinfo {author} {\bibfnamefont {E.}~\bibnamefont
  {Braaten}},\ }\bibfield  {title} {\bibinfo {title} {Harmonic and subharmonic
  association of universal dimers in a thermal gas},\ }\href
  {https://doi.org/10.1103/PhysRevA.92.013425} {\bibfield  {journal} {\bibinfo
  {journal} {Phys. Rev. A}\ }\textbf {\bibinfo {volume} {92}},\ \bibinfo
  {pages} {013425} (\bibinfo {year} {2015})}\BibitemShut {NoStop}%
\bibitem [{\citenamefont {Mark}\ \emph {et~al.}(2007)\citenamefont {Mark},
  \citenamefont {Ferlaino}, \citenamefont {Knoop}, \citenamefont {Danzl},
  \citenamefont {Kraemer}, \citenamefont {Chin}, \citenamefont {N\"{a}gerl},\
  and\ \citenamefont {Grimm}}]{Mark2007SOU}%
  \BibitemOpen
  \bibfield  {author} {\bibinfo {author} {\bibfnamefont {M.}~\bibnamefont
  {Mark}}, \bibinfo {author} {\bibfnamefont {F.}~\bibnamefont {Ferlaino}},
  \bibinfo {author} {\bibfnamefont {S.}~\bibnamefont {Knoop}}, \bibinfo
  {author} {\bibfnamefont {J.~G.}\ \bibnamefont {Danzl}}, \bibinfo {author}
  {\bibfnamefont {T.}~\bibnamefont {Kraemer}}, \bibinfo {author} {\bibfnamefont
  {C.}~\bibnamefont {Chin}}, \bibinfo {author} {\bibfnamefont {H.-C.}\
  \bibnamefont {N\"{a}gerl}},\ and\ \bibinfo {author} {\bibfnamefont
  {R.}~\bibnamefont {Grimm}},\ }\bibfield  {title} {\bibinfo {title}
  {Spectroscopy of ultracold trapped cesium {F}eshbach molecules},\ }\href
  {https://doi.org/10.1103/PhysRevA.76.042514} {\bibfield  {journal} {\bibinfo
  {journal} {Phys. Rev. A}\ }\textbf {\bibinfo {volume} {76}},\ \bibinfo
  {pages} {042514} (\bibinfo {year} {2007})}\BibitemShut {NoStop}%
\bibitem [{\citenamefont {Falco}\ and\ \citenamefont
  {Stoof}(2005)}]{Falco2005AMT}%
  \BibitemOpen
  \bibfield  {author} {\bibinfo {author} {\bibfnamefont {G.~M.}\ \bibnamefont
  {Falco}}\ and\ \bibinfo {author} {\bibfnamefont {H.~T.~C.}\ \bibnamefont
  {Stoof}},\ }\bibfield  {title} {\bibinfo {title} {Atom-molecule theory of
  broad {F}eshbach resonances},\ }\href
  {https://doi.org/10.1103/PhysRevA.71.063614} {\bibfield  {journal} {\bibinfo
  {journal} {Phys. Rev. A}\ }\textbf {\bibinfo {volume} {71}},\ \bibinfo
  {pages} {063614} (\bibinfo {year} {2005})}\BibitemShut {NoStop}%
\bibitem [{\citenamefont {Weber}\ \emph {et~al.}(2003)\citenamefont {Weber},
  \citenamefont {Herbig}, \citenamefont {Mark}, \citenamefont {N\"agerl},\ and\
  \citenamefont {Grimm}}]{Weber2003BEC}%
  \BibitemOpen
  \bibfield  {author} {\bibinfo {author} {\bibfnamefont {T.}~\bibnamefont
  {Weber}}, \bibinfo {author} {\bibfnamefont {J.}~\bibnamefont {Herbig}},
  \bibinfo {author} {\bibfnamefont {M.}~\bibnamefont {Mark}}, \bibinfo {author}
  {\bibfnamefont {H.-C.}\ \bibnamefont {N\"agerl}},\ and\ \bibinfo {author}
  {\bibfnamefont {R.}~\bibnamefont {Grimm}},\ }\bibfield  {title} {\bibinfo
  {title} {{Bose-Einstein} {C}ondensation of {C}esium},\ }\href
  {https://doi.org/10.1126/science.1079699} {\bibfield  {journal} {\bibinfo
  {journal} {Science}\ }\textbf {\bibinfo {volume} {299}},\ \bibinfo {pages}
  {232} (\bibinfo {year} {2003})}\BibitemShut {NoStop}%
\bibitem [{\citenamefont {Partridge}\ \emph {et~al.}(2005)\citenamefont
  {Partridge}, \citenamefont {Strecker}, \citenamefont {Kamar}, \citenamefont
  {Jack},\ and\ \citenamefont {Hulet}}]{Partridge2005MPO}%
  \BibitemOpen
  \bibfield  {author} {\bibinfo {author} {\bibfnamefont {G.~B.}\ \bibnamefont
  {Partridge}}, \bibinfo {author} {\bibfnamefont {K.~E.}\ \bibnamefont
  {Strecker}}, \bibinfo {author} {\bibfnamefont {R.~I.}\ \bibnamefont {Kamar}},
  \bibinfo {author} {\bibfnamefont {M.~W.}\ \bibnamefont {Jack}},\ and\
  \bibinfo {author} {\bibfnamefont {R.~G.}\ \bibnamefont {Hulet}},\ }\bibfield
  {title} {\bibinfo {title} {{Molecular Probe of Pairing in the BEC-BCS
  Crossover}},\ }\href {https://doi.org/10.1103/PhysRevLett.95.020404}
  {\bibfield  {journal} {\bibinfo  {journal} {Phys. Rev. Lett.}\ }\textbf
  {\bibinfo {volume} {95}},\ \bibinfo {pages} {020404} (\bibinfo {year}
  {2005})}\BibitemShut {NoStop}%
\bibitem [{\citenamefont {Chotia}\ \emph {et~al.}(2012)\citenamefont {Chotia},
  \citenamefont {Neyenhuis}, \citenamefont {Moses}, \citenamefont {Yan},
  \citenamefont {Covey}, \citenamefont {Foss-Feig}, \citenamefont {Rey},
  \citenamefont {Jin},\ and\ \citenamefont {Ye}}]{Chotia2012lld}%
  \BibitemOpen
  \bibfield  {author} {\bibinfo {author} {\bibfnamefont {A.}~\bibnamefont
  {Chotia}}, \bibinfo {author} {\bibfnamefont {B.}~\bibnamefont {Neyenhuis}},
  \bibinfo {author} {\bibfnamefont {S.~A.}\ \bibnamefont {Moses}}, \bibinfo
  {author} {\bibfnamefont {B.}~\bibnamefont {Yan}}, \bibinfo {author}
  {\bibfnamefont {J.~P.}\ \bibnamefont {Covey}}, \bibinfo {author}
  {\bibfnamefont {M.}~\bibnamefont {Foss-Feig}}, \bibinfo {author}
  {\bibfnamefont {A.~M.}\ \bibnamefont {Rey}}, \bibinfo {author} {\bibfnamefont
  {D.~S.}\ \bibnamefont {Jin}},\ and\ \bibinfo {author} {\bibfnamefont
  {J.}~\bibnamefont {Ye}},\ }\bibfield  {title} {\bibinfo {title} {{Long-Lived
  Dipolar Molecules and Feshbach Molecules in a 3D Optical Lattice}},\ }\href
  {https://doi.org/10.1103/PhysRevLett.108.080405} {\bibfield  {journal}
  {\bibinfo  {journal} {Phys. Rev. Lett.}\ }\textbf {\bibinfo {volume} {108}},\
  \bibinfo {pages} {080405} (\bibinfo {year} {2012})}\BibitemShut {NoStop}%
\bibitem [{\citenamefont {Zhang}\ \emph {et~al.}(2020)\citenamefont {Zhang},
  \citenamefont {Yu}, \citenamefont {Cairncross}, \citenamefont {Wang},
  \citenamefont {Picard}, \citenamefont {Hood}, \citenamefont {Lin},
  \citenamefont {Hutson},\ and\ \citenamefont {Ni}}]{Zhang2020FSM}%
  \BibitemOpen
  \bibfield  {author} {\bibinfo {author} {\bibfnamefont {J.~T.}\ \bibnamefont
  {Zhang}}, \bibinfo {author} {\bibfnamefont {Y.}~\bibnamefont {Yu}}, \bibinfo
  {author} {\bibfnamefont {W.~B.}\ \bibnamefont {Cairncross}}, \bibinfo
  {author} {\bibfnamefont {K.}~\bibnamefont {Wang}}, \bibinfo {author}
  {\bibfnamefont {L.~R.~B.}\ \bibnamefont {Picard}}, \bibinfo {author}
  {\bibfnamefont {J.~D.}\ \bibnamefont {Hood}}, \bibinfo {author}
  {\bibfnamefont {Y.-W.}\ \bibnamefont {Lin}}, \bibinfo {author} {\bibfnamefont
  {J.~M.}\ \bibnamefont {Hutson}},\ and\ \bibinfo {author} {\bibfnamefont
  {K.-K.}\ \bibnamefont {Ni}},\ }\bibfield  {title} {\bibinfo {title} {Forming
  a single molecule by magnetoassociation in an optical tweezer},\ }\href
  {https://doi.org/10.1103/PhysRevLett.124.253401} {\bibfield  {journal}
  {\bibinfo  {journal} {Phys. Rev. Lett.}\ }\textbf {\bibinfo {volume} {124}},\
  \bibinfo {pages} {253401} (\bibinfo {year} {2020})}\BibitemShut {NoStop}%
\bibitem [{\citenamefont {Spence}(2023)}]{Spence2023ASR}%
  \BibitemOpen
  \bibfield  {author} {\bibinfo {author} {\bibfnamefont {S.~J.}\ \bibnamefont
  {Spence}},\ }\emph {\bibinfo {title} {Assembling Single RbCs Molecules with
  Optical Tweezers}},\ \href {http://etheses.dur.ac.uk/14814/} {Ph.D. thesis},\
  \bibinfo  {school} {Durham University}, \bibinfo {address} {Durham
  University} (\bibinfo {year} {2023})\BibitemShut {NoStop}%
\bibitem [{\citenamefont {Petrov}\ \emph {et~al.}(2005)\citenamefont {Petrov},
  \citenamefont {Salomon},\ and\ \citenamefont {Shlyapnikov}}]{Petrov2005dmi}%
  \BibitemOpen
  \bibfield  {author} {\bibinfo {author} {\bibfnamefont {D.~S.}\ \bibnamefont
  {Petrov}}, \bibinfo {author} {\bibfnamefont {C.}~\bibnamefont {Salomon}},\
  and\ \bibinfo {author} {\bibfnamefont {G.~V.}\ \bibnamefont {Shlyapnikov}},\
  }\bibfield  {title} {\bibinfo {title} {Diatomic molecules in ultracold
  {F}ermi gases -- novel composite bosons},\ }\href
  {https://doi.org/10.1088/0953-4075/38/9/014} {\bibfield  {journal} {\bibinfo
  {journal} {J. Phys. B}\ }\textbf {\bibinfo {volume} {38}},\ \bibinfo {pages}
  {S645} (\bibinfo {year} {2005})}\BibitemShut {NoStop}%
\bibitem [{\citenamefont {Petrov}\ \emph {et~al.}(2004)\citenamefont {Petrov},
  \citenamefont {Salomon},\ and\ \citenamefont {Shlyapnikov}}]{Petrov2004wbd}%
  \BibitemOpen
  \bibfield  {author} {\bibinfo {author} {\bibfnamefont {D.~S.}\ \bibnamefont
  {Petrov}}, \bibinfo {author} {\bibfnamefont {C.}~\bibnamefont {Salomon}},\
  and\ \bibinfo {author} {\bibfnamefont {G.~V.}\ \bibnamefont {Shlyapnikov}},\
  }\bibfield  {title} {\bibinfo {title} {{Weakly Bound Dimers of Fermionic
  Atoms}},\ }\href {https://doi.org/10.1103/PhysRevLett.93.090404} {\bibfield
  {journal} {\bibinfo  {journal} {Phys. Rev. Lett.}\ }\textbf {\bibinfo
  {volume} {93}},\ \bibinfo {pages} {090404} (\bibinfo {year}
  {2004})}\BibitemShut {NoStop}%
\bibitem [{\citenamefont {Inguscio}\ \emph {et~al.}(2008)\citenamefont
  {Inguscio}, \citenamefont {Ketterle},\ and\ \citenamefont
  {Salomon}}]{Inguscio2008ufg}%
  \BibitemOpen
  \bibinfo {editor} {\bibfnamefont {M.}~\bibnamefont {Inguscio}}, \bibinfo
  {editor} {\bibfnamefont {W.}~\bibnamefont {Ketterle}},\ and\ \bibinfo
  {editor} {\bibfnamefont {C.}~\bibnamefont {Salomon}},\ eds.,\ \href@noop {}
  {\emph {\bibinfo {title} {Ultra-cold Fermi Gases, Proceedings of the
  International School of Physics "Enrico Fermi", Course CLXIV, Varenna, June
  2006}}}\ (\bibinfo  {publisher} {IOS Press, Amsterdam},\ \bibinfo {year}
  {2008})\BibitemShut {NoStop}%
\bibitem [{\citenamefont {Jag}\ \emph {et~al.}(2016)\citenamefont {Jag},
  \citenamefont {Cetina}, \citenamefont {Lous}, \citenamefont {Grimm},
  \citenamefont {Levinsen},\ and\ \citenamefont {Petrov}}]{Jag2016lof}%
  \BibitemOpen
  \bibfield  {author} {\bibinfo {author} {\bibfnamefont {M.}~\bibnamefont
  {Jag}}, \bibinfo {author} {\bibfnamefont {M.}~\bibnamefont {Cetina}},
  \bibinfo {author} {\bibfnamefont {R.~S.}\ \bibnamefont {Lous}}, \bibinfo
  {author} {\bibfnamefont {R.}~\bibnamefont {Grimm}}, \bibinfo {author}
  {\bibfnamefont {J.}~\bibnamefont {Levinsen}},\ and\ \bibinfo {author}
  {\bibfnamefont {D.~S.}\ \bibnamefont {Petrov}},\ }\bibfield  {title}
  {\bibinfo {title} {Lifetime of {F}eshbach dimers in a {F}ermi-{F}ermi mixture
  of $^{6}\text{Li}$ and $^{40}\text{K}$},\ }\href
  {https://doi.org/10.1103/PhysRevA.94.062706} {\bibfield  {journal} {\bibinfo
  {journal} {Phys. Rev. A}\ }\textbf {\bibinfo {volume} {94}},\ \bibinfo
  {pages} {062706} (\bibinfo {year} {2016})}\BibitemShut {NoStop}%
\bibitem [{\citenamefont {Thomas}\ and\ \citenamefont
  {Kj{\ae}rgaard}(2020)}]{Thomas2020ADF}%
  \BibitemOpen
  \bibfield  {author} {\bibinfo {author} {\bibfnamefont {R.}~\bibnamefont
  {Thomas}}\ and\ \bibinfo {author} {\bibfnamefont {N.}~\bibnamefont
  {Kj{\ae}rgaard}},\ }\bibfield  {title} {\bibinfo {title} {A digital feedback
  controller for stabilizing large electric currents to the ppm level for
  {F}eshbach resonance studies},\ }\href {https://doi.org/10.1063/1.5128935}
  {\bibfield  {journal} {\bibinfo  {journal} {Rev. Sci. Instrum.}\ }\textbf
  {\bibinfo {volume} {91}},\ \bibinfo {pages} {034705} (\bibinfo {year}
  {2020})}\BibitemShut {NoStop}%
\end{thebibliography}%

\end{document}